\pdfoutput=1 
\documentclass{JINST}
\newcommand{\ALD}{Atomic Layer Deposition~}

\newcommand{\ASIC}{Application-Specific Integrated Circuit~}
\newcommand{\DOE}{Department of Energy~}
\newcommand{\DUSEL}{Deep Underground Science and Engineering
Laboratory~}

\newcommand{\LAPPDTM}{LAPPD\textsuperscript{TM}~}
\newcommand{\LAPPD}{Large-Area Picosecond PhotoDetector~}
\newcommand{\KtwoNaSb}{K$_2$NaSb~}
\newcommand{\KtwoCsSb}{K$_2$CsSb~}

\newcommand{\UofC}{University of Chicago}
\newcommand{\UofCsp}{University of Chicago~}
\newcommand{\SLAC}{Stanford Linear Accelerator Laboratory~}
\newcommand{\SSL}{Space Sciences Laboratory~}
\newcommand{\SSLB}{Space Sciences Laboratory, University of California, Berkeley~}
\newcommand{\twentysquare}{20 $\times$ 20 cm$^2$~}
\newcommand{\fivesquare}{5 $\times$ 5 cm$^2$~}
\newcommand{\psec}{$10^{-12}$ s}

\usepackage{afterpage}
\title{A Brief Technical History of the Large-Area Picosecond
Photodetector (LAPPD) Collaboration}

\author{B. W. Adams$^a,$\footnote{Present Address:Incom, Inc. 294 
Southbridge Road, Charlton, MA 01507}~, 
K. Attenkofer$^{a,}$\footnote{Present Address:Brookhaven National
Laboratory, Upton NY, 11973}~, 
M. Bogdan$^c$,K. Byrum$^a$, A. Elagin$^c$, J. W. Elam$^a$,
H. J. Frisch$^c$, 
J.F.Genat$^c$, 
H. Grabas$^{c,}$\footnote{Present Address: Univ. of California Santa
Cruz, Santa Cruz CA 95064}~,
J. Gregar$^a$, E. Hahn$^d$, M.  Heintz$^c$,
Z. Insepov$^a$, V. Ivanov$^{f}$, S. Jelinsky$^g$, S. Jokela$^a$, S.
W. Lee$^a$, A. U. Mane$^a$, J. McPhate$^g$, M. J. Minot$^e$, P.  Murat$^d$,
K. Nishimura$^{h,}$\footnote{Present Address: Ultralytics LLC,
Arlington VA 22203}~,
R. Northrop$^c$, 
R.  Obaid$^{c,}$\footnote{Present Address: University of Connecticut, Storrs CT, 06269}~,
E. Oberla$^c$,
E. Ramberg$^d$, A. Ronzhin$^d$, O. H. Siegmund$^h$, G.  Sellberg$^d$,
N. T. Sullivan$^b$, A. Tremsin$^g$, G. Varner$^h$, 
I.  Veryovkin$^{a,}$\footnote{Present Address: University of Illinois, Chicago IL 60607}~,
A. Vostrikov$^c$, 
R. G. Wagner$^a$, D. Walters$^a$, H.-H. Wang$^a$, 
M. Wetstein$^{c,}$\footnote{Present Address: Iowa State University, Ames, IA 50011}~,
J. Xie$^a$, 
Z. Yusof$^{a,}$\footnote{Present Address: 
Illinois Institute of Technology, Chicago, IL 60616}~,
A. Zinovev$^a$\\
\llap{$^a$} Argonne National Laboratory, Argonne IL, 60439 \\
\llap{$^b$} Arradiance Inc., Sudbury MA, 01776\\
\llap{$^c$}Enrico Fermi Institute, University of Chicago, Chicago IL
60637 \\
\llap{$^d$} Fermi National Accelerator Laboratory, Batavia IL, 60510 \\
\llap{$^e$} Incom, Inc, Charlton MA 01507\\
\llap{$^f$} Muons,Inc;  Batavia IL 60610\\
\llap{$^g$}Space Sciences Laboratory, University of California,
Berkeley CA. 94720\\
\llap{$^h$} University of Hawaii, Honolulu HI, 96822\\
E-mail: \email{frisch@hep.uchicago.edu}}

\clearpage
\abstract{ The \LAPPD (LAPPD) Collaboration was formed in 2009 to
develop large-area photodetectors capable of time resolutions measured
in pico-seconds (psec, $10^{-12}$ s), with accompanying sub-millimeter
spatial resolution. During the next three and one-half years the
Collaboration developed the \LAPPDTM design of \twentysquare modules
with gains greater than $10^7$ and non-uniformity less than $15\%$,
time resolution less than 50 psec for single photons and spatial
resolution of 700~$\mu$m in both lateral dimensions.  We describe the
R\&D performed to develop large-area micro-channel plate glass
substrates, resistive and secondary-emitting coatings, large-area
bialkali photocathodes, and RF-capable hermetic packaging.  In
addition, the Collaboration developed the necessary electronics for
large systems capable of precise timing, built up from a custom
low-power 15-GigaSample/sec waveform sampling 6-channel integrated
circuit and supported by a two-level modular data acquisition system
based on Field-Programmable Gate Arrays for local control,
data-sparcification, and triggering. We discuss the formation,
organization, and technical successes and short-comings of the
Collaboration. The Collaboration ended in December 2012 with a
transition from R\&D to commercialization.}


\begin{document}

\section{Introduction}
\label{sec:introduction} The \LAPPD 
Collaboration (LAPPD) was formed in 2009 to develop photodetectors
capable of making time measurements measured in pico-seconds (psec,
\psec) with sub-mm spatial resolution over areas measured in
square-meters~\cite{LAPPD}. Micro-channel plates (MCPs) consist of
pores with the small spatial dimensions necessary for psec
timing~\cite{Credo, timing_paper,Wiza}. MCPs also naturally provide
homogeneity over large areas, excellent spatial resolution, high gain,
and low noise. Here we summarize the R\&D undertaken by LAPPD to
realize an MCP-based \twentysquare photodetector module with gain
greater than $10^7$, dark noise less than 0.1 counts/cm$^2$-s, time
resolution less than 50 psec for single photons, and spatial
resolution of 700 $\mu$m in each of the two lateral dimensions.  In
addition, LAPPD developed the necessary electronics for large systems
capable of precise timing, built up from a custom low-power
15-GigaSample/sec (GS/sec) waveform sampling 6-channel integrated
circuit and supported by a two-level modular data acquisition system
based on Field-Programmable Gate Arrays (FPGA) for local control,
data-sparcification, and triggering.

The project benefited from  the
availability of funding from the American Recovery and Reinvestment
Act (ARRA)~\cite{ARRA} through the \DOE (DOE). The ARRA funding, which
came to LAPPD via Argonne National Laboratory (ANL), allowed supporting
collaborating institutions with expertise in the needed technologies.
Strong support from the DOE based on the possibility of large
cost-savings at the proposed
\DUSEL (DUSEL)~\cite{DUSEL,DUNE} facility was crucial.

The LAPPD Collaboration was formed with participants from each of
three complementary kinds of institution, national laboratories,
universities, and US companies, each with its own strengths, culture,
and limitations. Industry brought specific manufacturing facilities
and industrial techniques; the national laboratories brought access to
a wide range of facilities and expertise for the synthesis and
characterization of materials; the universities brought specific
photodetector and electronics expertise. The project gained enormously
from having experts in many unanticipated areas of material science
and electronics, for example, who might not have become involved in a
more conventional collaboration.

The objectives of this paper are: 1) to provide a review of the major
technical accomplishments of LAPPD; 2) to provide guidance based on
experience for starting and sustaining a major joint
Laboratory-Industry-University effort;
and 3) to provide the historical context. The organization of the
paper is as follows.  The science motivation, early technical ideas,
and formation of the Collaboration are described in
Section~\ref{sec:beginnings}. The organization into five major
technical areas and the corresponding multi-disciplinary management
structure are described in Section~\ref{sec:organization}.  The
Collaboration goals, divided into those corresponding to each of the
five technical areas, are presented in Section~\ref{sec:goals}.
Section~\ref{sec:achievements} summarizes technical
R\&D. Section~\ref{sec:goals_assessment} assesses the achievement of
the goals, including the failure to make a fully-functional sealed
\LAPPDTM module before the transition to commercialization.
Section~\ref{sec:facilities} briefly describes the facilities that the
individual institutions constructed or otherwise brought to bear on
the development of capillary substrates, resistive and emissive
coatings, performance characterization, UHV-hermetic mechanical
packaging, and GHz electronics. The motivating role played by the joint
proposal from the DOE and the National Science Foundation (NSF) for a
large underground neutrino facility, the \DUSEL (DUSEL), is summarized
in Section~\ref{sec:dusel}.  Section~\ref{sec:communicating} presents
the infrastructure developed to communicate technical information in
many disciplines both within and external to the collaboration.
Section~\ref{sec:end_of_project} describes the end of the project in
Dec. 2012 and the subsequent transition to
commercialization~\cite{LAPPDTM}, with continuing R\&D under the DOE
detector and SBIR/STTR programs and private funding. The author list
and executive summary from the 2009 DOE proposal are shown in Appendix
A.

%
%
%
\section{Motivation and Selection of the Technology for Psec Timing}
\label{sec:beginnings} The need for very precise timing, with a
nominal goal of 1 psec, grew out of the recognition of the inability
to extract all the measurable information (i.e. 4-vectors) from
multi-TeV particle collisions at the Fermilab Tevatron in
2003~\cite{Aspen}. A detector capable of psec-resolution
timing\footnote{For convenience, we refer to the measurement of
time-of-arrival of photons in the deep sub-nsec range, i.e. from 10's
of psec to several hundred fsec, as `psec'. More information on
factors that determine the ultimate limit on timing resolution in
these technologies can be found in Ref~\cite{limitations_workshop}.}
must have physical dimensions small enough so that variations in the
fastest transit times of the photons or electrons forming the 
amplified signal correspond to a time jitter smaller than the desired
resolution~\cite{timing_paper}. Using Cherenkov light generated by
charged particles traversing a transparent radiator on the front face
of a Micro-Channel Plate was recognized in 2004 as a promising method
to cover large areas with sub-mm resolution~\cite{Credo}.

A second motivation was the suggestion by H.
Nicholson~\cite{Howard_Nicholson} that large-area inexpensive
panels of photodetectors could provide substantial cost savings by
replacing hundreds of millions of dollars of conventional
photomultipliers in the large water Cherenkov neutrino detector
proposed for DUSEL~\cite{DUSEL}. The number of these new
photodetectors needed for DUSEL was also financially attractive to companies
with existing facilities capable of producing large volumes of the
tubes, as described in Section~\ref{sec:dusel}.

Commercially available micro-channel plate
detectors~\cite{Photonis,Hamamatsu} had not been considered for
large-area applications due to their small size, with the largest
being \fivesquare~\cite{Planacon}; high-cost per area; long
fabrication cycle; and concerns about lifetime due to ion
feedback. Any new effort required making detectors with at least an
order-of-magnitude larger sensitive area at comparable or lower unit
cost, and a new process for fabricating the amplification section that
eliminated the source of the ion feedback. In addition, the supporting
electronics systems to exploit the fast timing capabilities for large
psec-capable systems needed to be defined and developed.
 
\subsection
{The Critical Roles of Seed Funding and Fermilab Test Beam Results}
\label{sec:seed_funding}

Seed funding was essential. The experimental program was started
in 2005 with a modest grant from the Physical Science Division at
the \UofC, followed by a three-year Laboratory Directed Research and
Development (LDRD) program at ANL~\cite{LDRD}, a two-year grant from
the joint ANL-Fermilab-Chicago program~\cite{joint_ANL_UC_FNAL}, and two
years of travel support from the Chicago-France Center of the
\UofCsp for US-France collaboration on the
electronics~\cite{Chicago_France}. The seed funding allowed a group
from ANL, Chicago, Fermilab, Hawaii, Saclay/IRFU, and the \SLAC (SLAC)
to test proof-of-principle in 2008 in a small experiment~\cite{T979}
at the Fermilab MTEST facility~\cite{MTEST} using commercial Planacon
MCP-PMTs~\cite{Planacon} and custom RF anodes~\cite{Tang_Naxos}. The
test confirmed the earlier work of Inami et al~\cite{Ohshima} that
Cherenkov light generated in the front window of an MCP-PMT easily
gave time resolutions below 10
psec~\cite{Ohshima,T979,Vavra_Fermilab,Anatoly}. The measured timing
resolution supported the case that precision time-of-flight
measurements could be made over large areas using micro-channel-plate
photodetectors with transmission-line readouts and custom
psec-resolution digitizing electronics, but would require significant
R\&D. This conclusion led to the formation of the Collaboration and
the development of a proposal to the DOE for a 3-year development
program.

\subsection{Proposal to the Office of High Energy Physics, Department
of Energy}
\label{sec:proposal}
The summer and fall of 2008 were devoted to defining the goals,
identifying the tasks and desired groups and individuals to
address them, and then on defining and writing a proposal to the
Office of High Energy Physics of the Department of Energy. 

The Collaboration was self-assembled from three kinds of institutions:
National Laboratories, US-based industry, and universities.  The list
of authors and institutions from the 2009 DOE proposal is given in Appendix
A.  Because the proposal was self-generated by the group and not
programmatically driven, it was based on a `bottom-up' intellectual
interest by individual participants in their own areas of expertise.  

The effort at the Argonne National Laboratory (ANL) involved six
Divisions each with its own expertise and facilities: High Energy
Physics, X-ray Sciences, Chemistry, Mathematics and Computer Science,
Energy Systems, and Material Science.  In addition, the
Physical Sciences and Engineering Directorate supported the ANL Glass
Shop. Fermilab provided expertise in photomultipliers, bialkali photocathodes,
micro-assembly of detectors and test facilities, and a facility for
the evaporation of large-area precision 
metal coatings,  as well as test-beam facilities;
SLAC (Stanford Linear Accelerator Center) provided deep expertise in
photodetectors, test beam measurements, and timing measurements at the
outset.

US industry also played an essential role.
 Arradiance~\cite{Arradiance} provided expertise in ALD-functionalized
 MCPs. Muons, Inc~\cite{Muons_Inc}, working closely with the
 Mathematics and Computer Science Division at ANL, supplied code and
 expertise in simulation of shower development in MCPs. Synkera
 Technologies~\cite{Synkera} applied their expertise in Anodic
 Aluminum Oxide (AAO) capillary substrates, working closely with the
 Material Science Division of ANL.

Researchers from four university groups were authors of the initial
proposal. The Space Science Laboratory at UC, Berkeley brought
extensive experience in the manufacture of MCP-based detectors, a
large UHV-capable facility for tube manufacture and characterization,
and expertise in electronics, testing, and end-use. The Enrico Fermi
Institute at the University of Chicago provided expertise in front-end
electronics and large electronics systems, mechanical design and
engineering, detector testing and characterization, computer support
for integration and documentation, and, working closely with personnel
at the Advanced Photon Source, the capability of testing
\LAPPDTM photodetectors with 
psec-time resolution.  The University of Hawaii
brought extensive experience with Application Specific Integrated
Circuit (ASIC), front-end electronics, and large systems design. The
group from the University of Washington brought expertise in
photodetectors and applications.

The project summary page from the proposal submission to the DOE in
2009 is also reproduced in Appendix A. The three areas singled out for
development are: 1) high quantum efficiency photocathode development;
2) use of
\ALD to apply resistive and emissive surface coatings to the walls of
the pores of the large area capillary plates to produce high-gain
low-noise electron amplification; and 3) the development of custom
fast low-power waveform sampling integrated circuits coupled to
transmission lines for readout of micro-channel-based
photomultipliers. The technical developments and associated published
papers and patents in each of these areas and others addressed during
LAPPD are discussed below in Section~\ref{sec:achievements}.

%
%
\section{Organization of the Collaboration}
\label{sec:organization} 

The formation of a multi-disciplinary
collaboration to develop a broad set of new detector technologies
crossing many areas of thin-film technology and  material science
presented an unusual challenge in the context of traditional High
Energy Physics (HEP) support. Experts in non-traditional HEP areas
were often self-selected by their interest in the problems in
their own field; the non-HEP Divisions at ANL in particular proved
to be a fertile ground for expertise and facilities. However key
expertise and facilities were also provided by the industrial
partners and the universities. We describe the project
organization below.

\subsection{Management}
The LAPPD project was managed through the ANL HEP Division, with
subcontracts to the industrial partners and universities. H. Frisch
was Spokesperson and held a joint appointment at ANL, reporting to
H. Weerts as Division Director. K. Byrum and Frisch were Co-PI's on
the proposal; R. G. Wagner was the Program Physicist.

\begin{figure}[tbp] 
\centering
\includegraphics[width=1.05\textwidth]{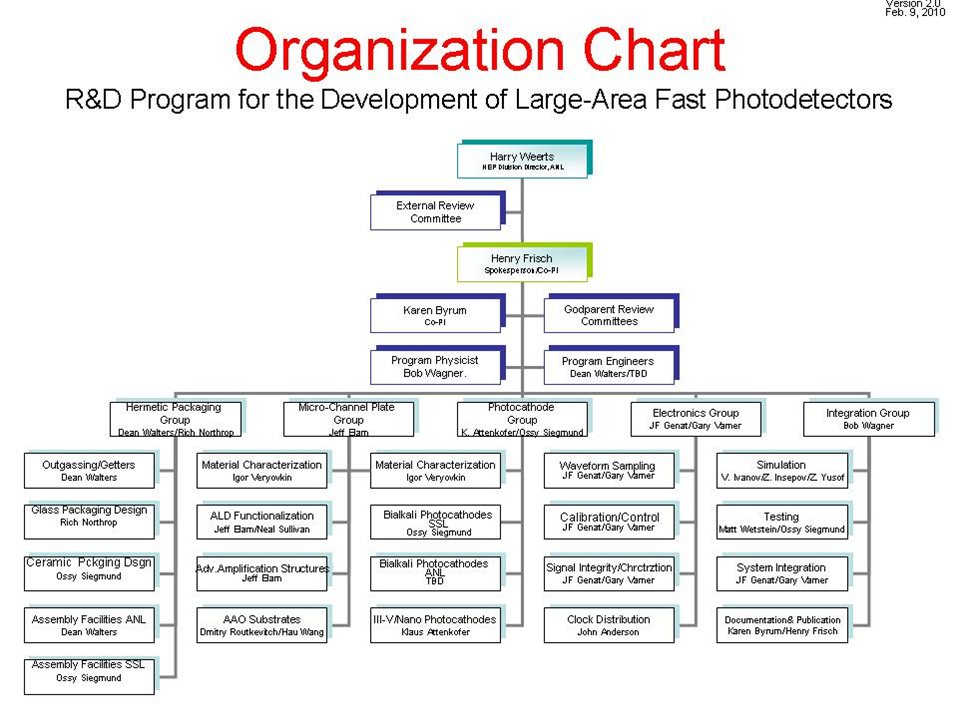}
\caption{The organization chart for the LAPPD Collaboration as of Feb.
2010. The effort was organized into five areas: Hermetic Packaging,
Micro-channel plates, Photocathodes, Electronics, and
Integration. Twice-yearly 'Godparent' Review Committee meetings and
Collaboration Meetings and a weekly `All-collaboration' meeting provided
the necessary communication between the groups.}
\label{fig:org_chart}
\end{figure}

The structure of the collaboration is shown graphically in
Figure~\ref{fig:org_chart}. The photodetector development
tasks were organized as four parallel
structures, with co-leaders from the appropriate areas of
expertise. Integration was identified as a separate
parallel task, as shown. These five areas are described in more
detail below in Section~\ref{sec:four_areas}.

In order to reduce the risk inherent in a project that had so many
new technologies, the hermetic packaging task was split into two
separate solutions, an evolutionary ceramic package by the  \SSL
 (SSL) group of O. Siegmund, and a  `frugal' all-glass package by
Chicago, ANL, and collaborators, as described in 
Section~\ref{sec:portfolio_of_risk}.

Communication of information across the many areas of technical
development in the collaboration was provided by bi-annual
`Godparent reviews' and Collaboration Meetings, a web blog and
library of figures, technical specifications (prints), and
documents, and a weekly meeting using the blog for agenda and
talks, as described in Section~\ref{sec:communicating}.

\subsection{Photodetector Module Development}
\label{sec:four_areas} The primary photodetector R\&D was divided 
up into the four areas shown in Figure~\ref{fig:org_chart}. Hermetic
Packaging refers to the development of a thin, robust, economical
sealed tube housing that can maintain ultra-high vacuum for
decades. The Micro-channel Plate Group was responsible for the
development of robust large-area capillary substrates and the
resistive and secondary-emitting coatings necessary to convert them
into high-gain low-noise MCPs. The Electronics/Integration Group was
responsible for developing digitizing electronics capable of psec
resolution, scalable multi-channel systems, and the interface to the
electrical and mechanical aspects of the package. The Photocathode
Group was formed to transfer photocathode fabrication techniques to
the large flat-panel format specific to LAPPD. In addition to these
four narrowly-focused groups, the Integration Group was responsible
for the Collaboration-wide tasks of simulation, testing, system
integration, and documentation.

The major necessary R\&D developments in each area
are listed below in Section~\ref{sec:goals}.

\subsection{Parallel Development of Ceramic and Glass Designs: "The Portfolio of Risk"}
\label{sec:portfolio_of_risk}
\begin{figure}[!t] 
\centering
\includegraphics[width=.85\textwidth]{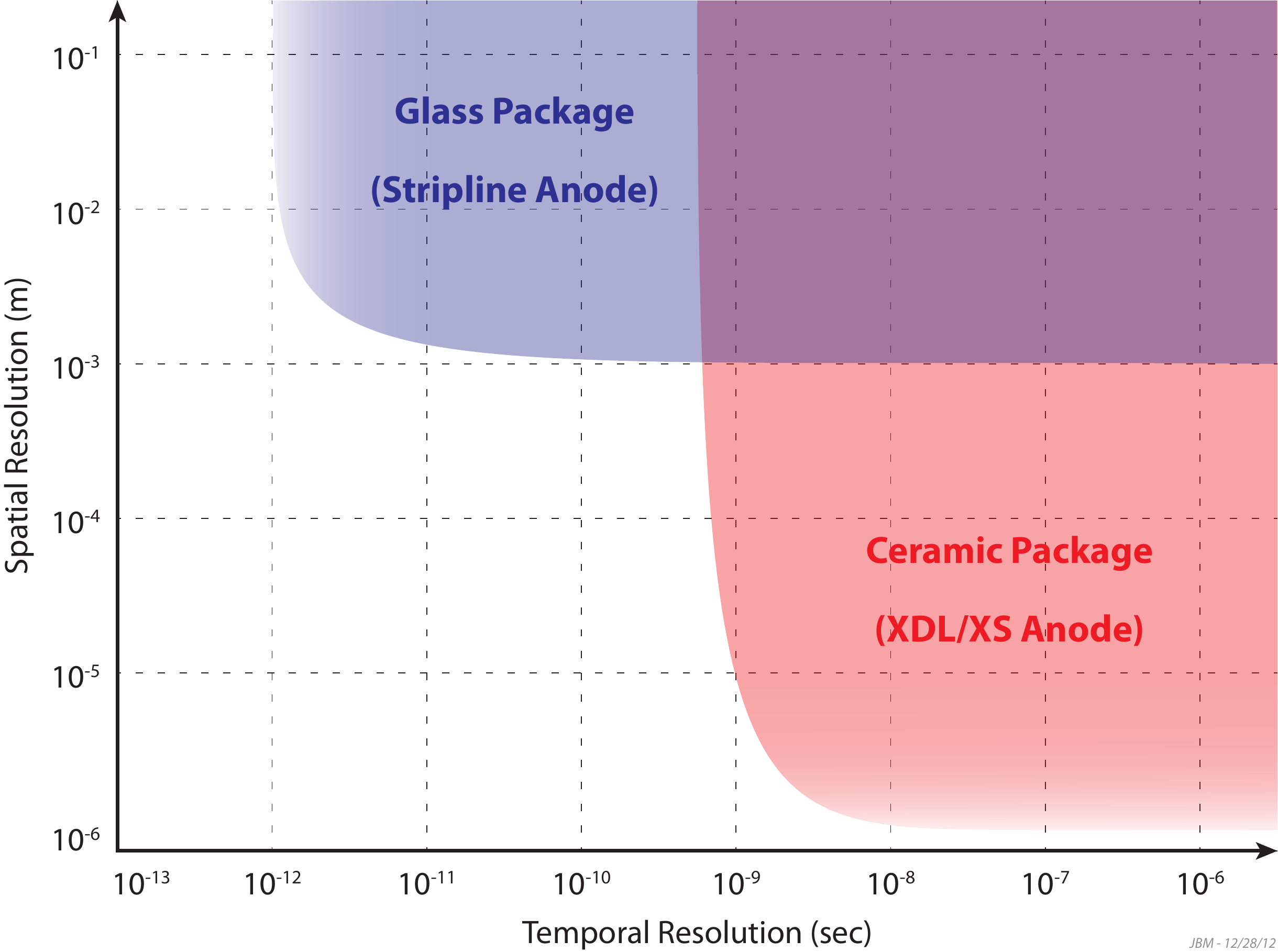}
\caption{The contrasting performance optimizations in time and space
resolution of the ceramic and glass implementations of the \LAPPDTM
module. The Micro-Channel Plates that form the amplification section of
both detectors are identical; the difference comes from choices made
for the resolution of the anode that collects the charge. The anode of
the glass package consists of high bandwidth micro-strip transmission
lines, optimized for time resolution; the finer-detailed anode of the
ceramic package has been optimized for spatial resolution. (credit:
J. McPhate)}
\label{fig:jason_ceramic_glass}
\end{figure}

 The Collaboration adopted a two-pronged approach to the hermetic
package, a ceramic package design, inspired by smaller MCP-PMT
designs, and a more radical, all-glass design. The original motivation
was risk mitigation, with the ``conventional'' ceramic module perceived
as being lower risk, but more expensive, and the novel glass module
design viewed as higher risk, but ultimately with fewer parts and a
simpler assembly.  The parallel efforts
allowed a coordinated program to develop resistive and emissive
coatings at the ANL facility with continuous access to MCP
testing at \SSL  and ANL/Chicago, and a flow of expertise from SSL
on the highly specialized handling and treatment of MCP-PMT
components and test equipment. As the two paths became better established
the complementary performance characteristics in space and time
resolution emerged, as shown in
Figure~\ref{fig:jason_ceramic_glass}.

\section{Goals}
\label{sec:goals}
At the time of the proposal the basic issues such as choice of
capillary substrate, feasibility of resistive and emissive coatings
for such large areas, feasibility of large systems of psec
electronics, and many others, were wide open. For example, although
we knew we needed a small pore size for fast timing, both glass
capillary and Anodic Aluminum Oxide substrates seemed viable
options. Whether one could uniformly coat large substrates
containing large aspect-ratio capillary pores was an open
question. The feasibility of building an electronics
system capable of maintaining psec resolution over large areas while
limiting the power consumption and channel count was a subject of debate.

Basic questions such as these informed the formation of the
Collaboration and the organization of the R\&D. We list below the
initial goals for each of the areas in
Tables~\ref{goals_MCPs}-\ref{goals_electronics}. For each goal we have
tabulated the LAPPD papers published describing the results of the
R\&D. A discussion of the successes and failures is given in
Section~\ref{sec:goals_assessment}.

\newpage
\begin{table}[tpb]
\centering
\begin{tabular}{|p{5.0in}|p{1.0in}|}
\hline\hline
\multicolumn{2}{|c|} {\bf\Large Goals: Microchannel Plates}\\
\hline 
{\bf ~~~~~~~~~~~~~~~~~~~~~~~~~~~Goal} & {\bf References} \\
\hline

Suitability (uniformity, open-area, manufacturability) of drawn glass
 capillary substrates as MCP's & \cite{Craven_TIPP,Minot_NIM}\\
 & \\

Suitability (uniformity, open-area, manufacturability)
of AAO etched aluminum substrates as MCPs & \cite{AAO_patent}\\
& \\

Down-selection decision between glass and AAO substrates & NA\\
& \\

Development of Atomic Layer Deposition Resistive coatings &  
\cite{Jeff_ALD_ECS_2013,Anil_ALD_ECS_2014,ALD_patent}\\
& \\

Development of Atomic Layer Deposition secondary-emitting
coatings & \cite{Slade_SEY_NIM,Slade_SEY_TIPP2011, Slade_SEY_TIPP2011,
 Insepov_SEY_simulation_comparison, Insepov_Ivanov_SEY, 
Baryshev_Igor_Jeff_profiling}\\
& \\

Development of high-yield manufacturing techniques for 8"
glass substrates & ~\cite{Craven_TIPP,Minot_NIM}\\
& \\

Implementation of facilities for characterization
of the 8" \LAPPDTM MCPs for robustness, gain, uniformity, life-time,
and time and position resolution & \cite{RSI_paper},
\cite{Ossy_JINST_2013}\\
\hline
\end{tabular}
\caption{The goals (Column 1) and published references (Column 2) 
for achievements of the LAPPD Microchannel Plate Group. The reader is
encouraged to access the references in the Bibliography
for technical details. In the Reference column we have
indicated achievements not suitable for publication in a peer-reviewed
journal by the abbreviation 'NA'; details for these are available from the
Collaboration Library web pages or by request.}
\label{goals_MCPs}
\end{table}
\newpage

\begin{table}[!tbp]
\centering
\begin{tabular}{|p{5.0in}|p{1.0in}|}
\hline\hline
\multicolumn{2}{|c|} {\bf\Large Goals: Photocathodes}\\
\hline 
{\bf ~~~~~~~~~~~~~~~~~~~~~~~~~~~Goal} & {\bf References} \\
\hline

Transfer of \SSL techniques for \KtwoNaSb
 photocathodes to 8"-square photocathodes on borosilicate glass & \cite{Ossy_JINST_2013}\\
& \\

Development/acquisition of  equipment and transfer of
techniques for \KtwoCsSb photocathodes to 8"-square photocathodes on
 borosilicate glass & ~\cite{Junqi_TIPP}\\
& \\

Establishing collaborative efforts within and outside of the
Collaboration to develop `theory-based' photocathodes with
reproducible quantum efficiency  near the predicted limit & \cite{RMD_SBIR,Luca_1,Luca_2}\\
& \\

\hline
\end{tabular}
\caption{The goals (Column 1) and published references (Column 2) 
for achievements of the LAPPD Photocathode Group.}
\label{goals_photocathode}
\end{table}

\newpage
\begin{table}[p]
\centering
\begin{tabular}{|p{5.0in}|p{1.0in}|}
\hline\hline
\multicolumn{2}{|c|} {\bf\Large Goals: Hermetic Packaging}\\
\hline 
{\bf ~~~~~~~~~~~~~~~~~~~~~~~~~~~Goal} & {\bf References} \\
\hline

Design of a `frugal' glass package using widely-available
float glass & \cite{HV_paper}\\
& \\

Design of a ceramic package evolved from the widely-used \SSL
designs for smaller MCP-PMT packages & \cite{Ossy_JINST_2013}\\
& \\

Development and characterization of inexpensive GHz
microstrip anodes with good (few psec) time resolution for the
glass package & \cite{anode_paper}\\
& \\

Development and characterization of high spatial resolution
(sub-mm) microstrip anodes for the ceramic package & \cite{Ossy_JINST_2013}\\
& \\

Development of the anode seal over the microstrips for the
glass package & \cite{Craven_TIPP,Minot_NIM,HV_paper}\\
& \\

Development of electrical contacts and pin structure for the
ceramic package & \cite{Ossy_JINST_2013}\\
& \\

Finite-element analysis of mechanical stresses to determine
anode and window thicknesses and support structure layout & NA\\
& \\

Development of the window seal for the glass package &
\cite{Craven_TIPP,Minot_NIM}\\ 
& \\

Development of the window seal for the ceramic package & \cite{Ossy_JINST_2013}\\
& \\

Identification of reliable and committed vendors for parts
and subsystems & NA \\
& \\

Production of a hermetic sealed \LAPPDTM module with photocathode &  See \S~\ref{sec:sealed_tile}\\

\hline
\end{tabular}
\caption{The goals (Column 1) and published references (Column 2) 
for achievements of the LAPPD Hermetic Packing Group. We note that the
goal of a top seal for the glass package has now been demonstrated.
In the References column the abbreviation 'NA' indicates developments
not appropriate for peer-reviewed publication.}
\label{goals_hermetic_packaging}
\end{table}

\newpage
\begin{table}[p]
\centering
\begin{tabular}{|p{5.0in}|p{1.0in}|}
\hline\hline
\multicolumn{2}{|c|} {\bf\Large Goals: Electronics and System Integration}\\
\hline 
{\bf ~~~~~~~~~~~~~~~~~~~~~~~~~~~Goal} & {\bf References} \\
\hline

Development and characterization of a multi-channel CMOS
\ASIC (ASIC) capable of wave-form sampling at greater than 10
GS/sec & \cite{PSEC4_paper}\\
& \\

Development of a corresponding scalable electronics system
with multi-mode triggering, FPGA local control and readout,  and a
simple low-rate generic DAQ interface & \cite{OTPC_paper}\\
& \\

Development of a large-area economical multi-module 'Supermodule'
using GHz-bandwidth anodes, fast waveform sampling, robust
mechanical packaging, and an optical-fiber
readout & \cite{anode_paper}\\
& \\

Development of test stands for full system testing and
characterization & \cite{RSI_paper}\\
& \\

Characterization, including gain, uniformity, lifetime, and
time and space resolutions, of the performance of the glass and
ceramic packages & \cite{Ossy_IEEE_2011,Ossy_JINST_2013,timing_paper}\\

\hline
\end{tabular}
\caption{The goals (Column 1) and published references (Column 2) 
for achievements of the LAPPD Electronics and Integration Groups.} 
\label{goals_electronics}
\end{table}
%

\newpage
\clearpage

%
%
\section{R\&D Achievements}
\label{sec:achievements}

The LAPPD goals were set out in annual milestones. All the milestones were
achieved, with the (glaring) exception of the production of a
self-standing hermetically sealed \LAPPDTM module, which only occurred
after the Collaboration had ended (see
Section~\ref{sec:goals_assessment}). Here we list some of the essential
R\&D outcomes.

\subsection{Microchannel Plates: Substrates and Coatings}

\begin{figure}[b!] 
\centering
\includegraphics[height=.30\textheight]{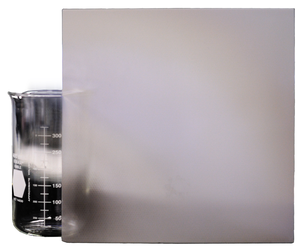} 
\hfil
\includegraphics[height=.30\textheight]{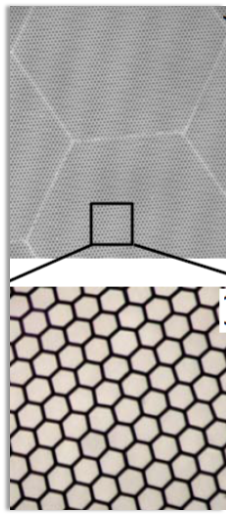}
\caption{Left: An $8"\times8"$ 
glass capillary substrate developed by the Incom corporation. Each
substrate contains approximately 80 million 20-micron-diameter
capillary pores. The open-area ratio is typically greater than
65\%, making the plate appear transparent. Right: A close-up of a
similar glass substrate showing the 20-micron capillary 
structure.} \label{fig:Incom_substrate}
\end{figure}

The Collaboration started with two alternatives for the large-area
microchannel-plate substrate, anodic aluminum oxide (AAO), and
glass. The AAO program was ended in a down- selection to conserve
resources, with a patent issued~\cite{AAO_patent}. For the glass
substrate, Incom~\cite{Incom} undertook the development, drawing hollow glass tubes
down to an inner diameter of 20 microns, assembling and fusing them
into a solid block large enough to provide 8"-square plates, and
cutting them on an 8$^o$ bias into 1.2 mm-thick wafers to produce a
ratio of pore length to diameter of 60:1. The substrates were then
manufactured to size and surface finish.  The seemingly simple tasks of
cleaning adequately for subsequent coating and then characterizing the
capillary substrates, each of which has a surface area of 6.5 m$^2$,
required extensive development before being successfully incorporated
into the production process. Figure~\ref{fig:Incom_substrate} shows an
Incom 8'' substrate; the transparency is due to an open-area ratio
exceeding 65\%.

\begin{figure}[btp] 
\centering
\includegraphics[width=.65\textwidth]{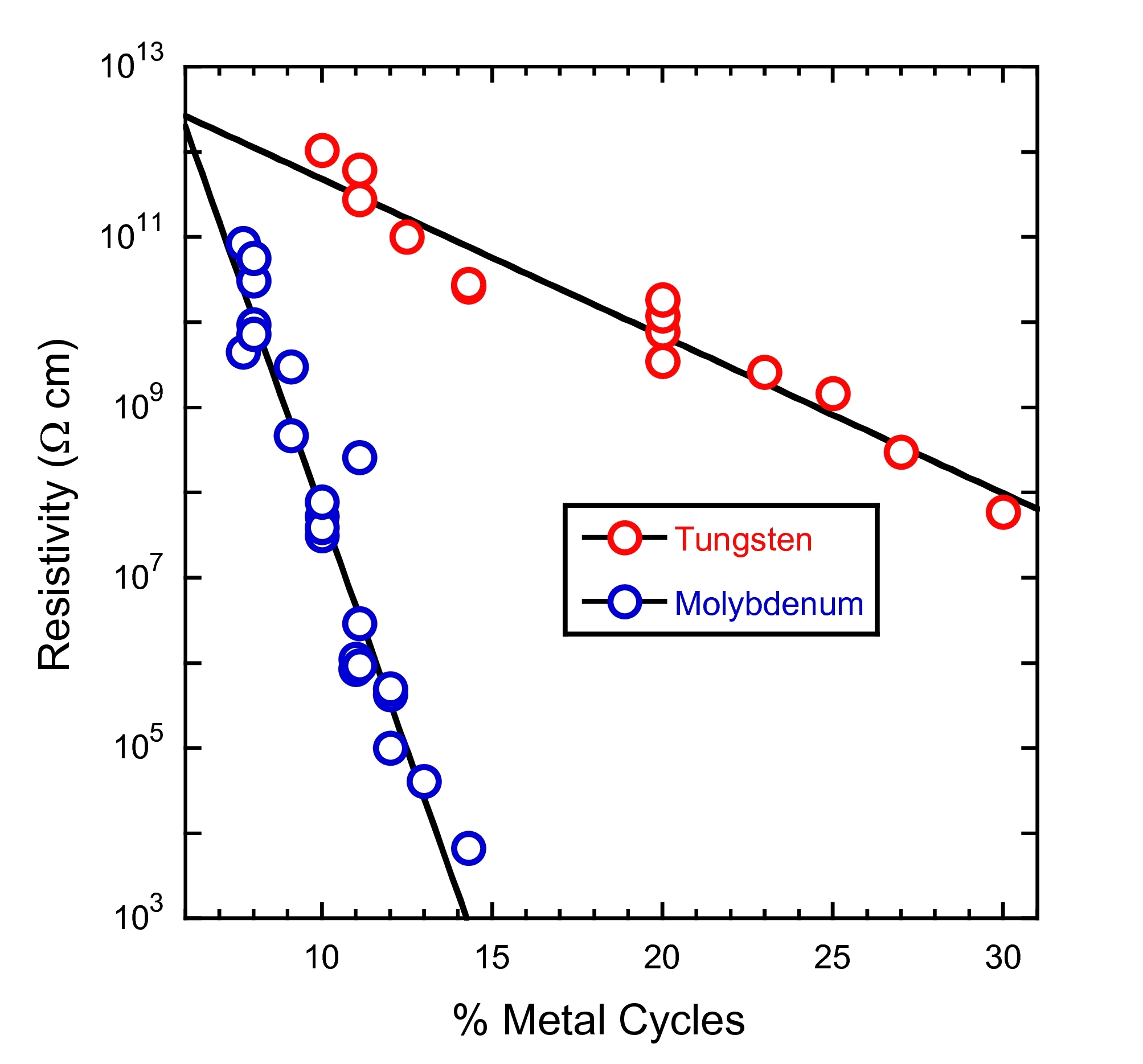}
\caption{Resistivity of the Argonne ALD tunable resistance coatings used for
functionalizing capillary glass arrays to produce MCPs versus the
number of metal cycles.  The resistivity of the molybdenum films has an
exponential slope versus the metal cycle percentage twice that of the
tungsten films because the amount of Mo deposited in a single Mo ALD cycle is
approximately twice that of the W
ALD~\cite{Jeff_ALD_ECS_2013,Anil_ALD_ECS_2014}.}
\label{fig:jeffe_ALD_resistivity}
\end{figure}

The use of ALD to coat passive substrates was demonstrated by
Arradiance before the start of LAPPD~\cite{Arradiance_coatings}, and
was instrumental in the LAPPD interest in ALD for coating large-area
arrays.  Figure~\ref{fig:jeffe_ALD_resistivity} shows the resistivity
of ALD tunable resistance coatings developed in the Energy Systems
Division at Argonne.  These coatings comprise conducting metallic
nano-particles (tungsten (W) or molybdenum (Mo)) in an amorphous
dielectric matrix (Al$_2$O$_3$).  The resistivity drops exponentially
with metal content, and the metal content is controlled by adjusting
the percentage of ALD metal cycles during the deposition of the
coatings.  The plot shows the exponential behavior versus the number
of metal cycles, a mixed blessing in that the behavior allows covering
a large range, but also has a high sensitivity to the metal content.
The resistivity of the molybdenum films is more sensitive to the metal
cycle percentage than that of the tungsten films because the amount of
Mo deposited in a single Mo ALD cycle is approximately twice that of
the W ALD~\cite{Jeff_ALD_ECS_2013,Anil_ALD_ECS_2014}

\begin{figure}[btp] 
\centering
\includegraphics[width=.65\textwidth]{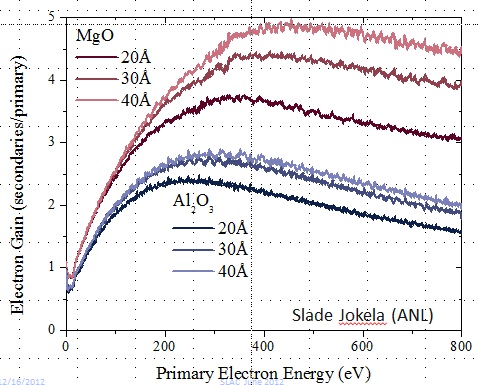}
\caption{ Measurements of the secondary emission of MgO and
Al$_2$O$_3$ versus incident electron energy~\cite{Slade_SEY_NIM,Slade_SEY_TIPP2011}.} 
\label{fig:slade_SEY}
\end{figure}

The emission of secondary electrons versus electron energy is an
 essential input into simulations of the cascade in the capillary
 pores, used to predict the voltage needed, gain, and pulse
 behavior~\cite{Valentin_NIM,Insepov_Ivanov_SEY}. A dedicated facility
 for the measurement of secondary emission yield (SEY) vs incident
 electron energy was constructed in the Materials Science Division
 (MSD) at ANL.  Figure~\ref{fig:slade_SEY} shows the measurements of
 SEY for two standard ALD thin film coatings, MgO and
 Al$_2$O$_3$~\cite{Slade_SEY_NIM,Slade_SEY_TIPP2011,
 Insepov_SEY_simulation_comparison}. In addition to higher gain, a
 higher SEY contributes to a narrower transit time distribution due to
 smaller fluctuations in yield in the first strike (the initiation of
 the shower by the photoelectron from the cathode).

\begin{figure}[b] 
\centering
\includegraphics[width=0.45\textwidth]{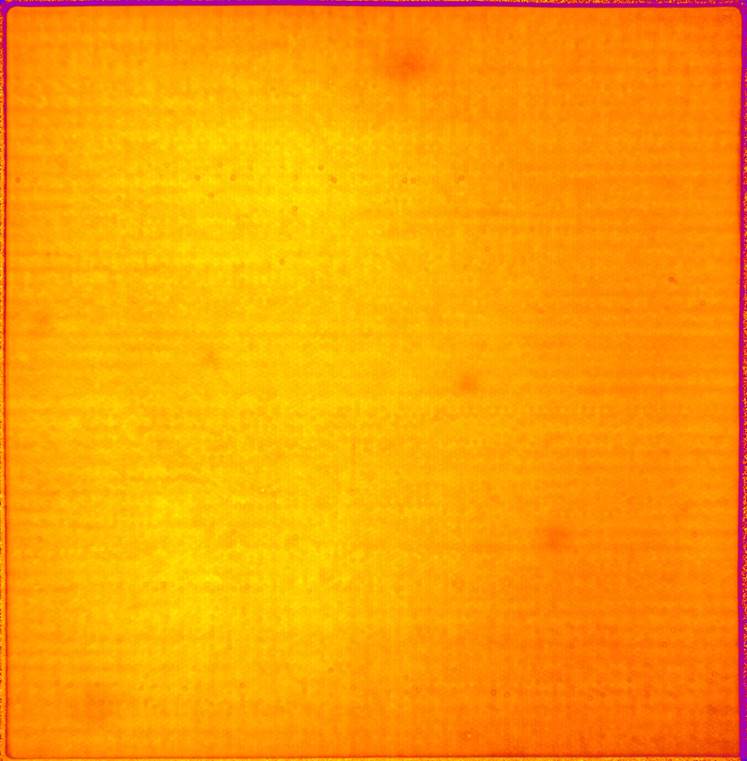}
\hfil
\includegraphics[width=0.45\textwidth]{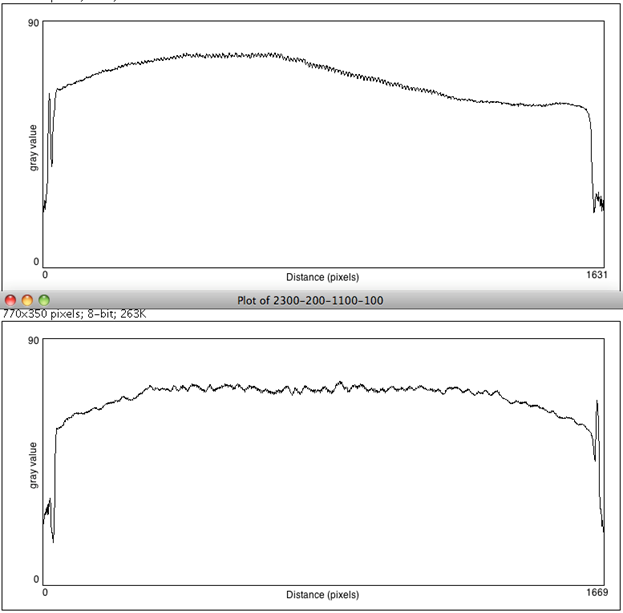}
 \caption{ Left: A gain map measured at SSL of a pair of fully
 functionalized 20-cm-square capillary plates. The mean gain of the
 pair is $7 \times 10^{6}$.  Right: The one-dimensional projections show a
 gain uniformity within the specification ($\pm 15$\%) adequate for
 track reconstruction in neutrino
 events~\cite{Ossy_IEEE_2011,Ossy_JINST_2013}.}
 \label{fig:ossy_anil_MCP_gain_uniformity}
\end{figure}

An image from the SSL test facility that measures gain and uniformity
over the full area of a pair of ALD-functionalized LAPPD MCP plates is
shown in Figure~\ref{fig:ossy_anil_MCP_gain_uniformity}. The
gain uniformity is within the $\pm 15\%$ requirement.

\begin{figure}[t] 
\centering
\includegraphics[width=.45\textwidth]{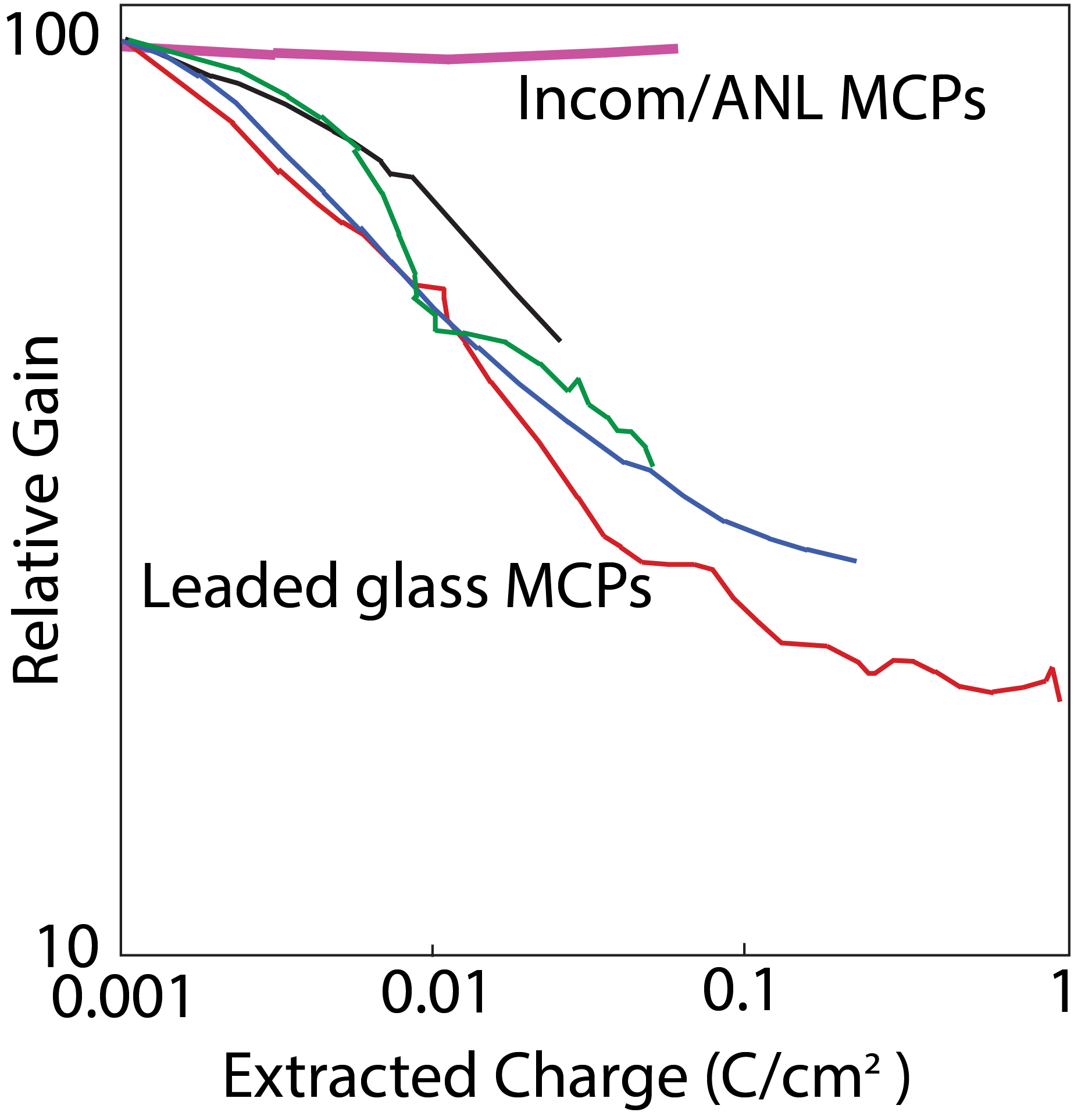}
\caption{Measurements at the \SSL of the stability versus charge
extracted of an ALD-coated MCP from the ANL ESD group compared to
commercial plates~\cite{{Ossy_IEEE_2011}}.}
\label{fig:ossy_lifetest}
\end{figure}

Figure~\ref{fig:ossy_lifetest} shows SSL measurements of the stability
of the ALD-coated MCP plates compared to conventional lead glass
plates versus the amount of charge extracted. The conventional plates
show the loss with `scrubbing', a long
(and hence expensive) burn-in process of drawing current from the plates
during production to achieve a
quasi-stability~\cite{Ossy_IEEE_2011,Ossy_JINST_2013}.

\subsection{Hermetic Packaging}
The  complementary packaging efforts in ceramic and glass
are described below in Section~\ref{sec:SSL_package}  and
Section~\ref{sec:UC_package}, respectively.

\subsubsection{The Brazed Ceramic Package}
\label{sec:SSL_package}
In order to stay close to their own proven techniques, the SSL group
developed a package design that was a direct evolution from their long
string of successful designs using well-tried techniques, in
particular a brazed high-purity ceramic-metal package, a captured amplification
section of MCP plates and spacers that could be electrically tested before the
window was sealed, and an indium-bismuth low temperature capillary
pressure window seal.

\begin{figure}[b]
\centering
\includegraphics[angle=0,height=.15\textheight]{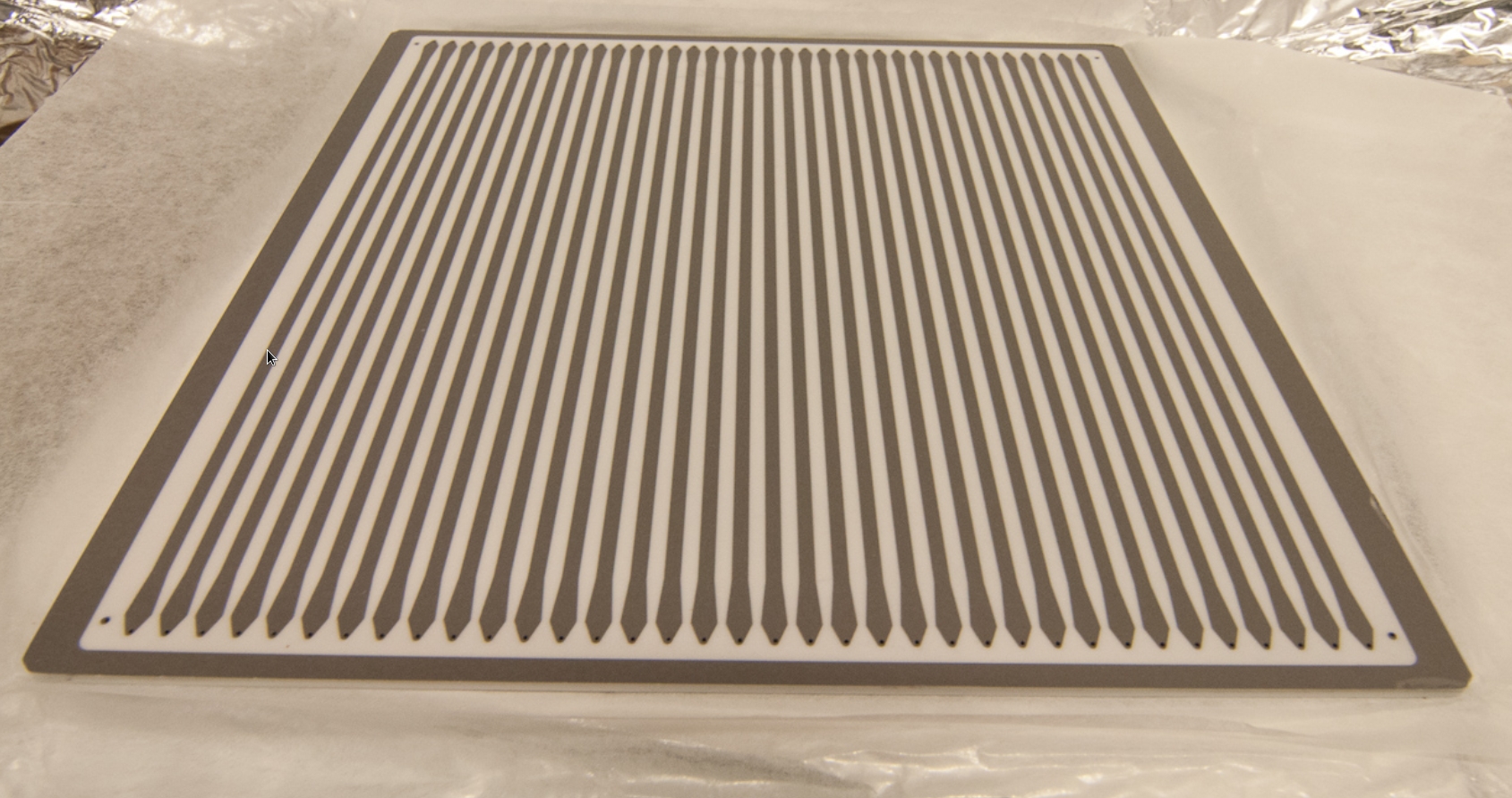}
\hfil
\includegraphics[angle=0,height=.15\textheight]{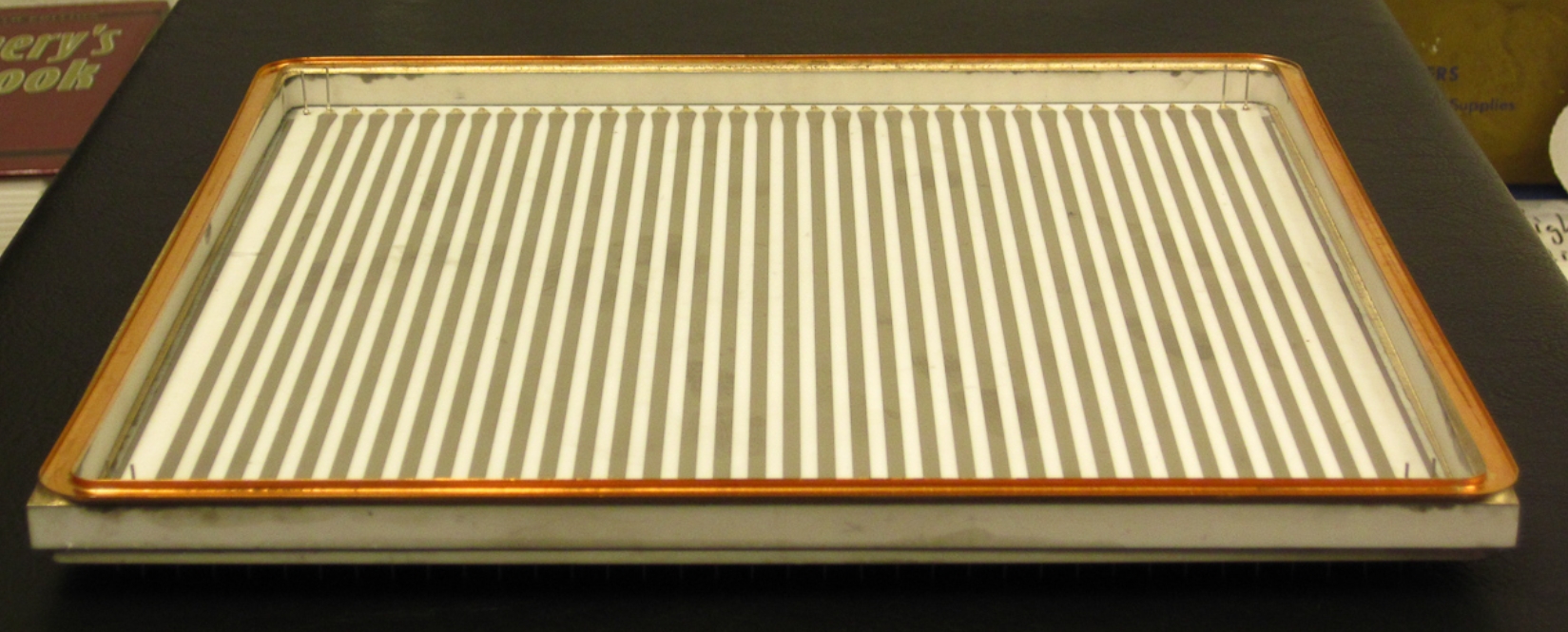}
\caption{Left: the Space Sciences Lab ceramic anode; Right: a complete
SSL module base assembly of anode, sidewall, and copper well for the
molten In-Bi solder seal.}
\label{fig:jason_anodes}
\end{figure}

The ceramic anode that forms the bottom surface of the SSL module is
shown in the left-hand panel of Figure~\ref{fig:jason_anodes}. The
right-hand panel shows the lower module assembly, consisting of the
anode brazed to a ceramic sidewall with a copper well on top that holds
the indium-bismuth eutectic that forms the low temperature seal to the
window.

\begin{figure}[b] 
\centering
\includegraphics[height=.26\textheight]{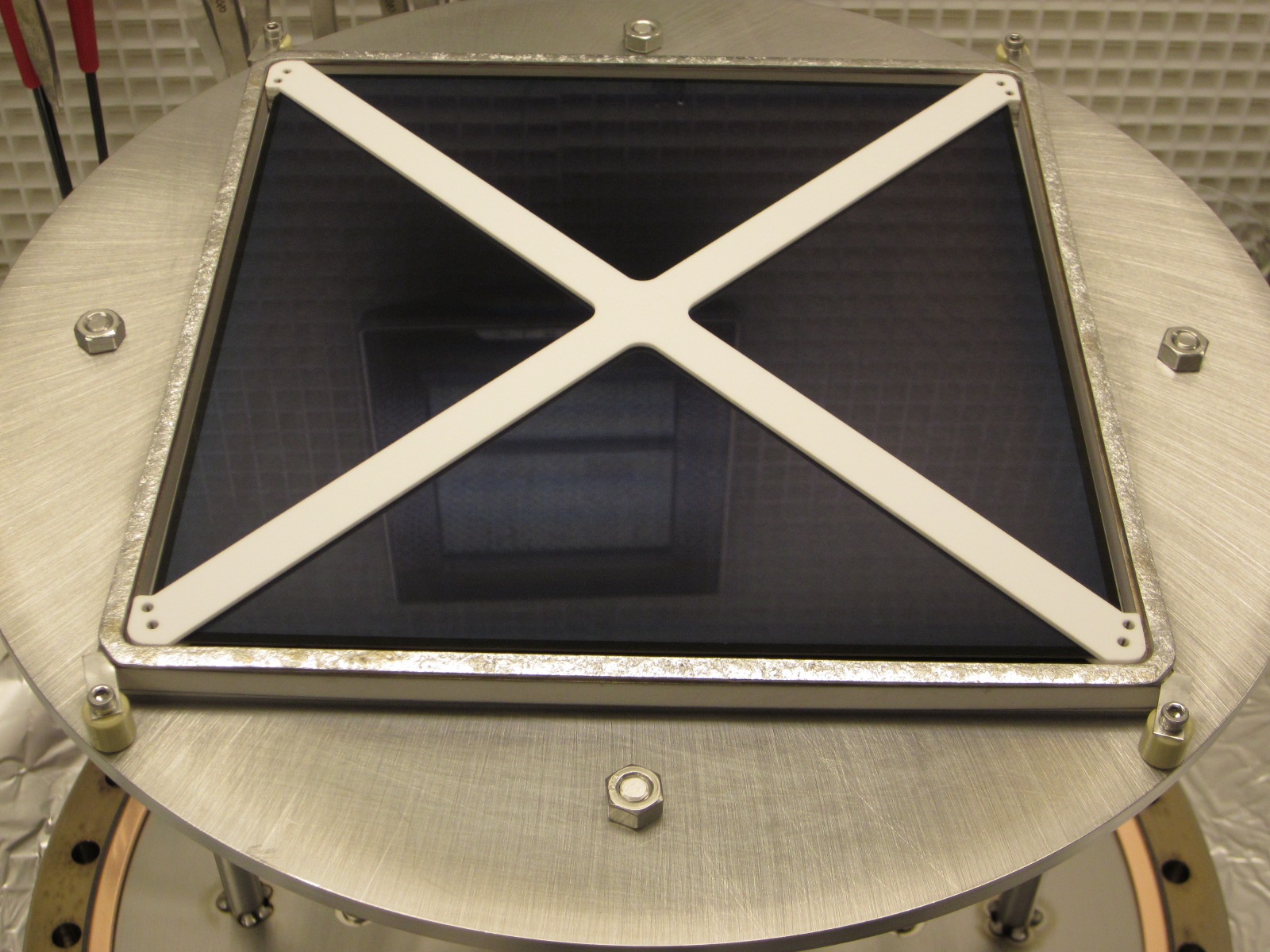}
\hfil
\includegraphics[height=.26\textheight]{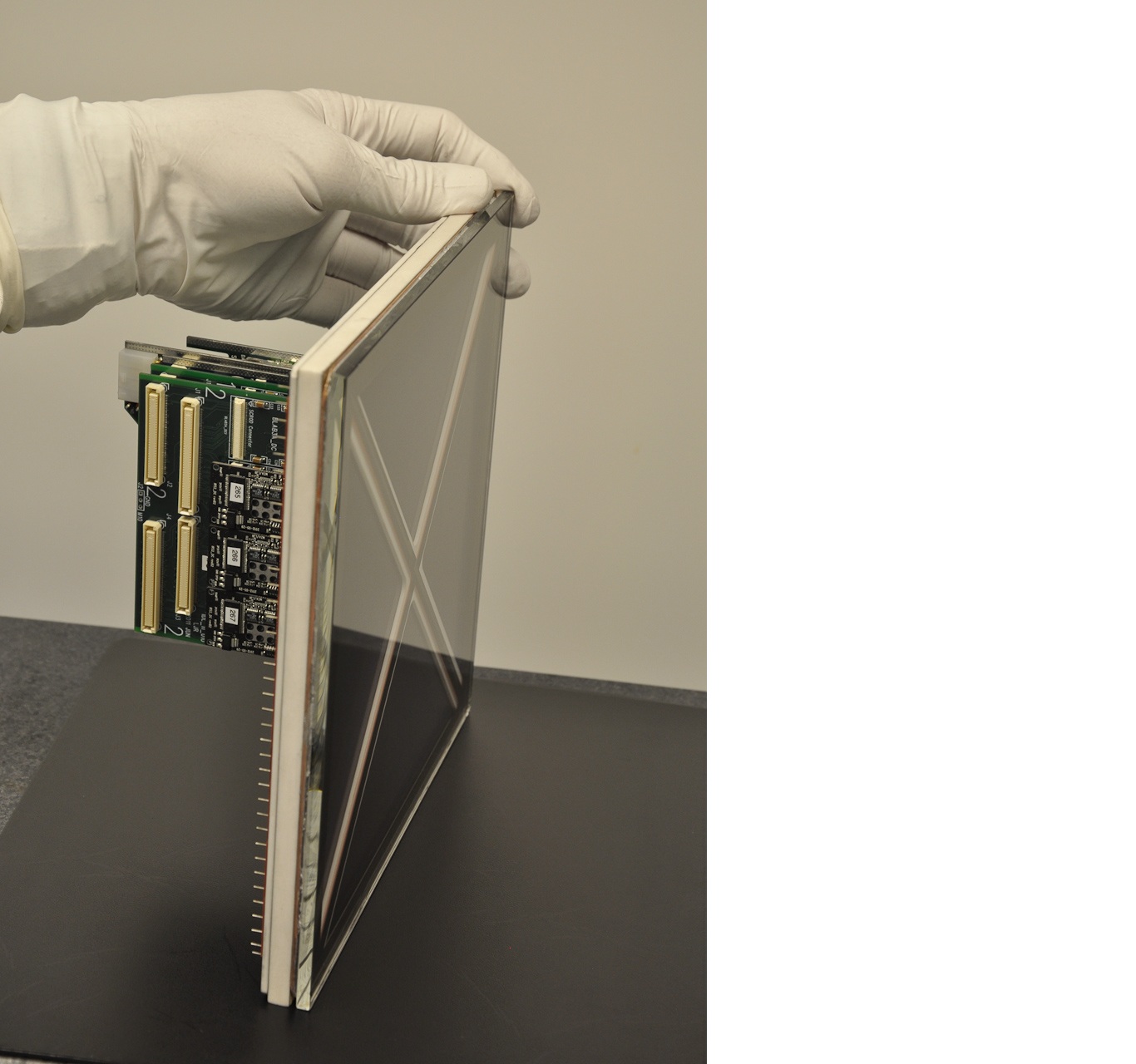}
\caption{Left:  The ceramic module base
                    prior to loading into the large SSL tank for photocathode
                    deposition and sealing of the window. The black
                    surface is the ALD coating on the top MCP. The 
		    top `X' spacer is also visible.
Right: The ceramic module package with the University of
                    Hawaii electronics attached to the anode pins on
                    the back. The SSL module uses a Na$_2$KSb cathode, a
                    spacer design consisting of a cross with an
                    embedded getter, and an anode design optimized for
                    position resolution. }
\label{fig:ceramic_package}
\end{figure}

Figure~\ref{fig:ceramic_package} shows the mechanical design of the 
full SSL module, including the
University of Hawaii electronics package that connects to the anode
pins on the back. 

A test of the fully-assembled SSL \LAPPDTM module with photocathode 
shortly after the Collaboration ended is described in 
Section~\ref{sec:sealed_tile}.


\subsubsection{The Glass  Hermetic Package}
\label{sec:UC_package}

\begin{figure}[btp] 
\centering
\includegraphics[width=0.50\textwidth]{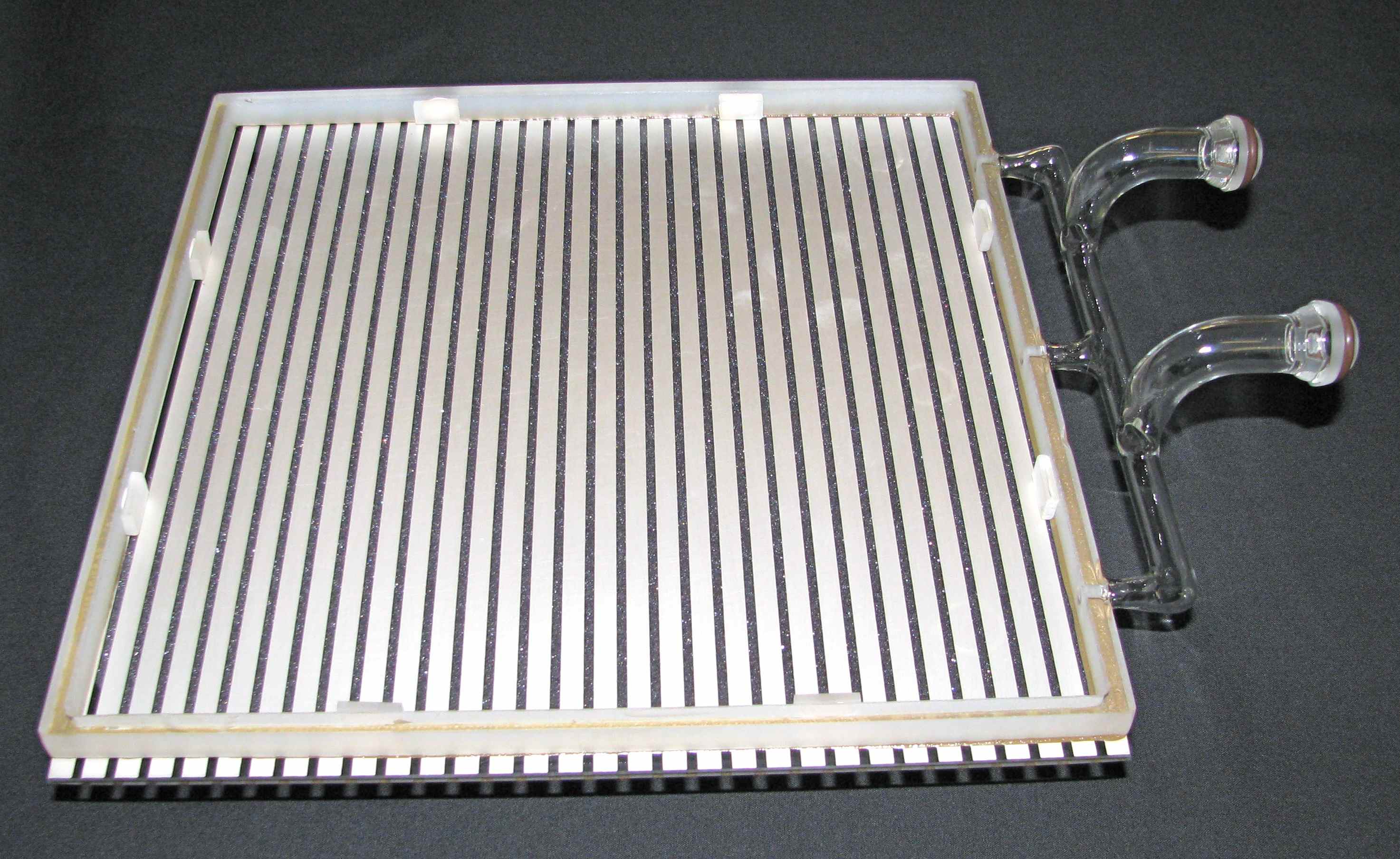}
\caption{A glass module base assembly showing the microstrip anode, and the
glass frit joint between the glass sidewall and the anode.  The two
ports on this assembly are specific to the Demountable test facility
at the Argonne Advanced Photon Source, in which the module is sealed
with an O-ring and the cathode is a thin aluminum layer. (credit:
Joseph Gregar, ANL Glass Shop)} 
\label{fig:joe_demountable_ltba}
\end{figure}

The glass module design was driven by the goal of achieving time
resolutions in the psec range. The mechanical design considerations
were consequently inseparable from the high-frequency (multi-GHz
signals) requirements; bringing these high-frequency signals out of
the glass vacuum package was identified very early as a challenge.
Figure~\ref{fig:joe_demountable_ltba} shows the solution, demonstrated
at Minotech~\cite{Minotech} and at the ANL Glass Shop. The microstrips
on the anode are run under the glass sidewall, and the hermetic seal
between the sidewall and the anode is made with a glass frit selected
for a match in thermal expansion with borosilicate glass.


\begin{figure}[btp] 
\centering
\includegraphics[height=0.30\textwidth]{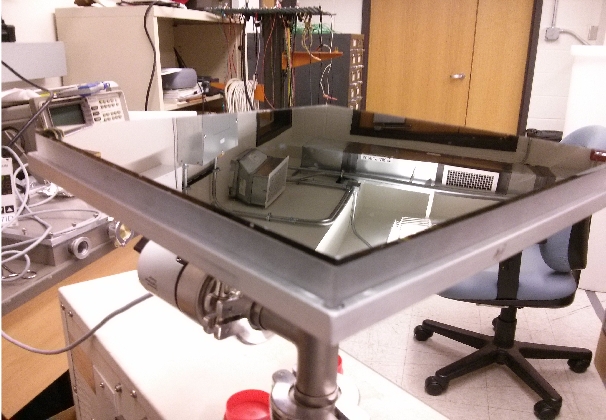}
\hfil
\includegraphics[height=0.30\textwidth]{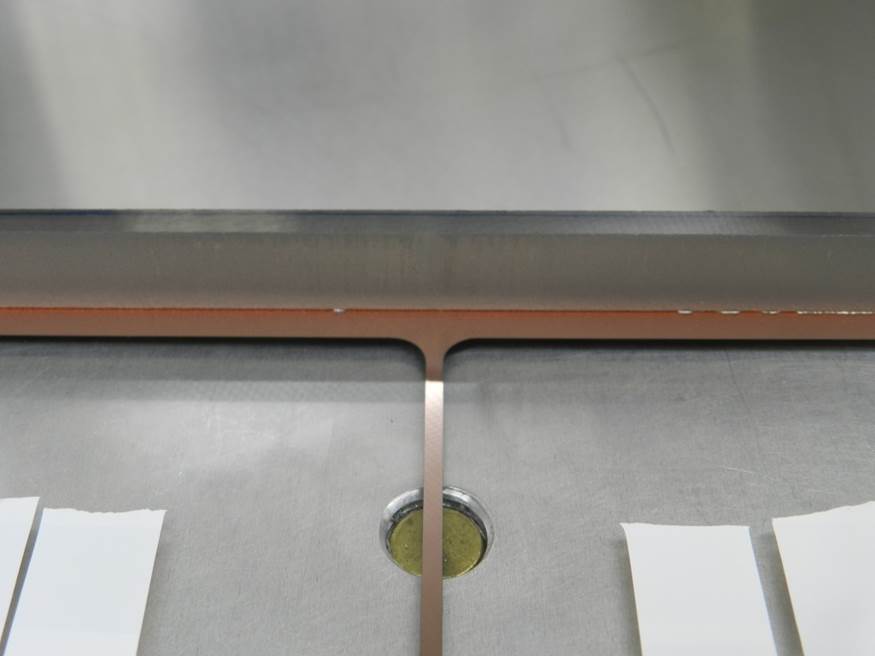}
\caption{A glass \LAPPDTM package with a hermetic indium 'flat' seal
between the glass sidewall and the glass top window
under test. Right: A close-up of the indium flat seal, taken through
a top window with no photocathode. The photo shows the
NiCr metalization on the top surface of the sidewall; a Cu layer is
visible where the indium has not taken it into solution. The Cu/NiCr
''finger'' extending towards the bottom of the frame is for distribution of
current towards the center of the cathode.}
\label{fig:andrey_topseal}
\end{figure}

The left-hand panel of Figure~\ref{fig:andrey_topseal} shows a
window, in this case with an aluminum cathode, 
hermetically sealed to the glass sidewall. The right-hand panel shows
a closeup of the hermetic seal between the window and the
sidewall. This seal is 'flat' (i.e. has no well) 
and uses pure indium instead of the In-Bi alloy.

\clearpage
\newpage

\subsection{Photocathodes}

\begin{figure}[tbh] 
\centering
\includegraphics[width=0.50\textwidth]{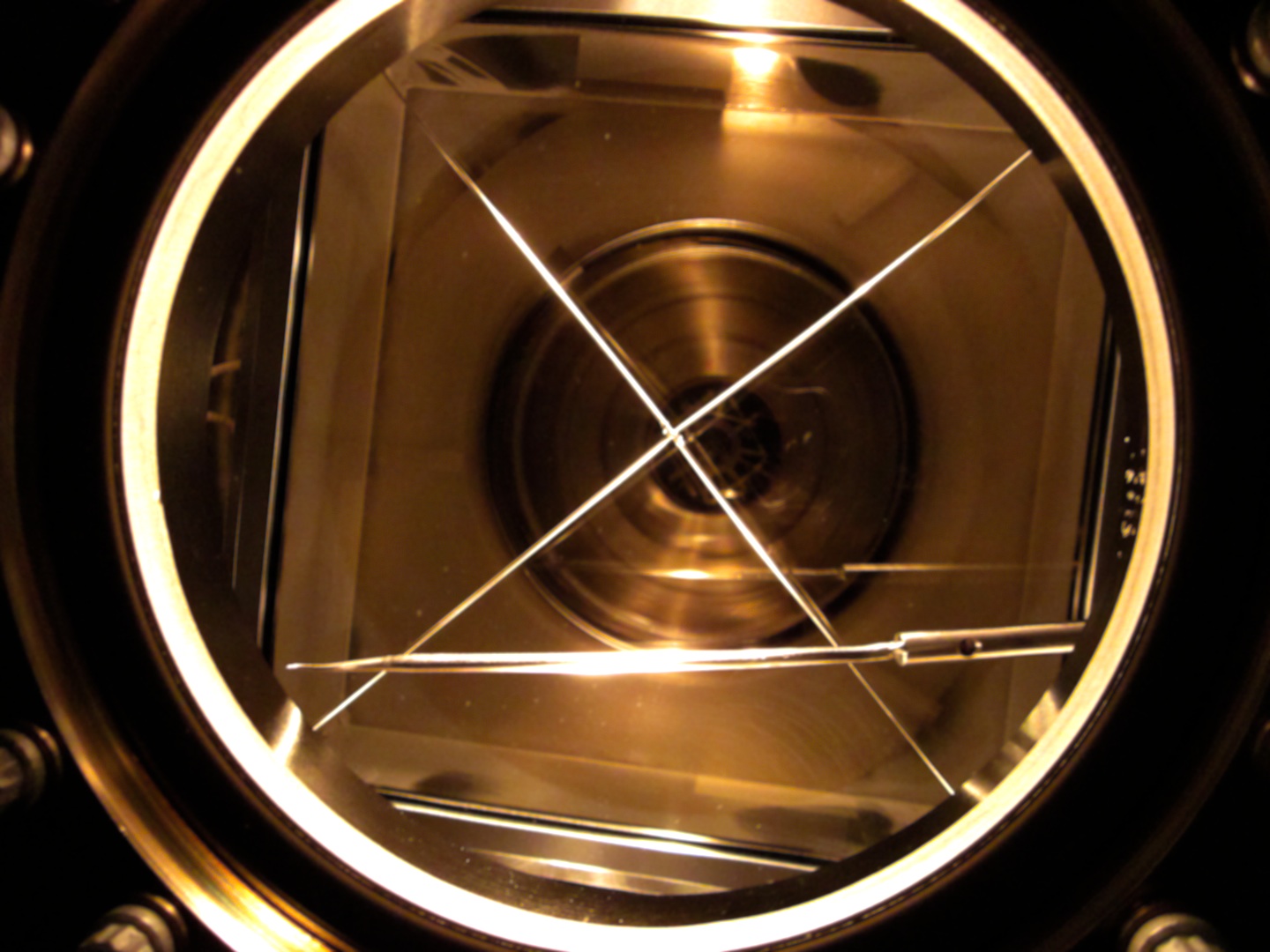}
 \caption{A \KtwoNaSb photocathode synthesized on an LAPPD window in
the large fabrication tank at the 
Space Sciences Laboratory~\cite{Ossy_IEEE_2011,Ossy_JINST_2013}.}
\label{fig:ossy_first_cathode}
\end{figure}

\begin{figure}[b] 
\centering
\includegraphics[width=0.50\textwidth]{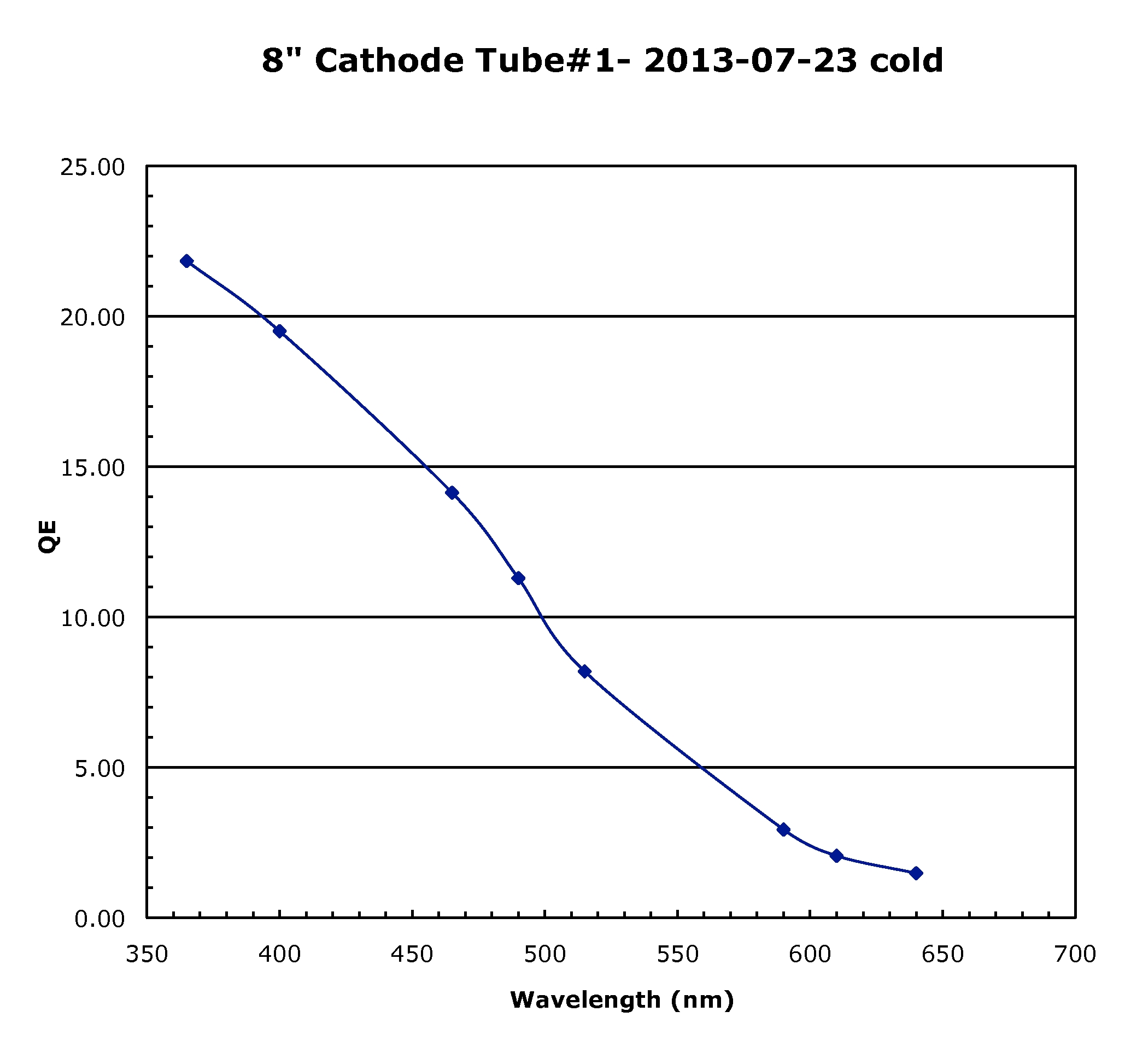}
 \caption{The quantum efficiency versus wavelength of an  8" \KtwoNaSb  
photocathode made in the large-tank facility at the \SSL on 
a B33 glass substrate~\cite{Ossy_JINST_2013}.}
\label{fig:ossy_QE}
\end{figure}

Figure~\ref{fig:ossy_first_cathode} shows the first full-size
photocathode made in the large-tank facility at the Space Sciences
Laboratory.  The SSL cathodes are chosen to be \KtwoNaSb for
temperature stability, good conductance, and low background.
Figure~\ref{fig:ossy_QE} shows the quantum efficiency versus
wavelength of an 8" \KtwoNaSb photocathode made in the large-tank
facility at SSL on a B33 glass substrate~\cite{Ossy_JINST_2013}.

\clearpage
\newpage

\subsection{Electronics}

The electronics for the ceramic package was based on 
the University of Hawaii waveform sampling ASICs~\cite{Hawaii_electronics}. 
Figure~\ref{fig:ceramic_package} shows the
University of Hawaii electronics package integrated with the anode
pins on the back of the ceramic module.

\begin{figure}[!h]
\centering
\includegraphics[angle=0,width=0.25\textwidth]{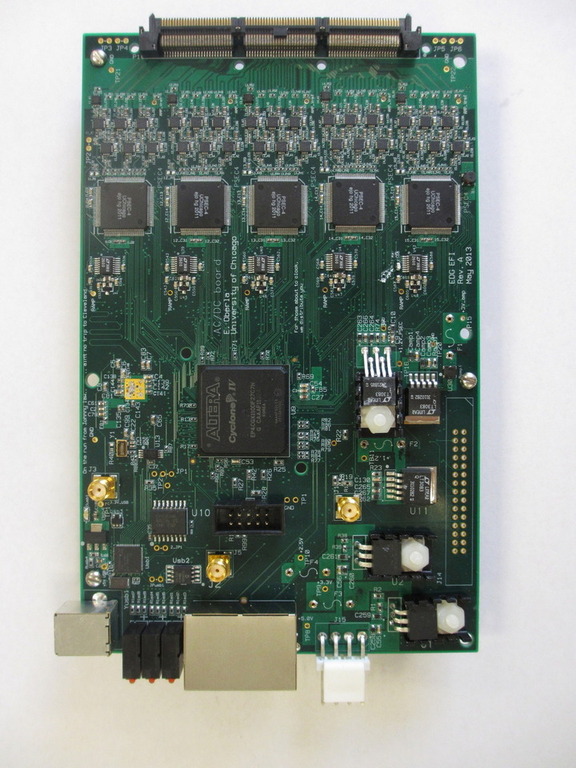}
\hfil
\includegraphics[angle=0,width=0.45\textwidth]{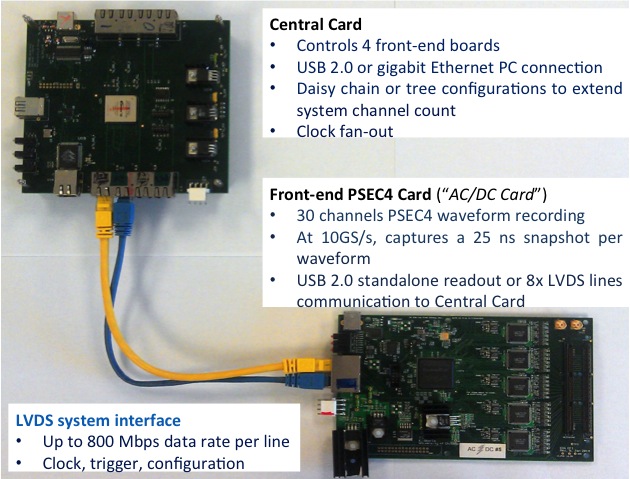}
\caption{Left: The ACDC front-end card, with five 6-channel 15
 GigaSample/sec PSEC4 digitizing chips. Right: A Central Card, used to
 read out and control up to eight ACDC cards, showing the connection to
 a single ACDC card. The system is now scalable up to 1920 channels
 per Central Card in a modular Master/Slave configuration.}
\label{fig:electronics}
\end{figure}
\begin{figure}[tbp] 
\centering
\includegraphics[width=.45\textwidth]{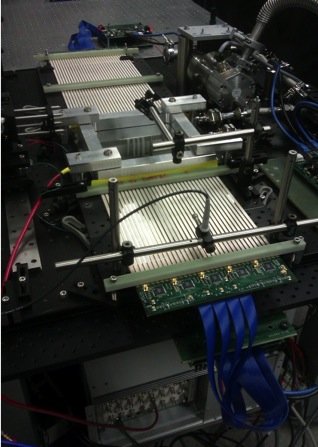}
\caption{The `Demountable' LAPPD module, 
mounted on a 30-strip anode that is 90 cm
long to accommodate 4 \LAPPDTM modules~\cite{anode_paper}.  The active
module is read out on both ends of the 4-module anode using
60-channels of PSEC4 waveform sampling and data
acquisition~\cite{PSEC4_paper}. Three such assemblies would form the
Supermodule.}
\label{fig:APS_Demountable}
\end{figure}

A custom integrated circuit, the PSEC4 chip~\cite{PSEC4_paper}, and
 supporting electronics, capable of time resolution measured in psec
 and scalable to large systems, were developed in parallel with the
 photodetector systems.  A front-end printed
 circuit card, the ACDC card, containing five 6-channel PSEC4 chips is
 shown in the left-hand panel of Figure~\ref{fig:electronics}. Up to eight
 of the 30-channel ACDC cards can be controlled by the Central Card,
 shown in the right-hand panel. A system of two ACDC cards and one
 Central Card, i.e. 60 channels, was used in the Demountable test
 setup, as shown in Figure~\ref{fig:APS_Demountable}.

\begin{figure}[!h]
\centering
\includegraphics[angle=0,width=0.45\textwidth]{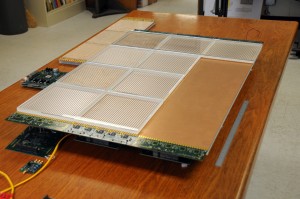}
\hfil
\includegraphics[angle=0,width=0.45\textwidth]{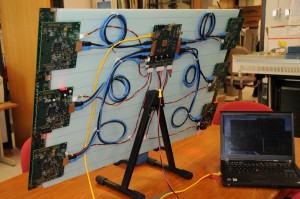}
\caption{Left: The design of an integrated system of 12 \LAPPDTM
modules consisting of three 4-module panels, each read-out on the ends by
ACDC cards. Right: A possible integration of the ACDC front-end
electronics and Central Card system on the 12-module system. 
The modularity of the anode-strip readout and electronics allows
application-specific variations on this basic structure, with
trade-offs in area coverage, bandwidth, and cost.} 
\label{fig:supermodule}
\end{figure}

For the glass module, the integration of electronics readout with a
large-area mechanical design (the `Supermodule') is illustrated in
Figures~\ref{fig:electronics} and ~\ref{fig:supermodule}. GigaHz
signals require an integrated treatment of mechanical packaging and
signal acquisition. The \LAPPDTM glass package design is modular,
allowing serial connection of the anode strips from one unit to the
next, with readout on the ends of the package. Measurements of
bandwidth and crosstalk were made with up to four
modules~\cite{anode_paper}. 

At the APS lab at Argonne~\cite{RSI_paper}, a full detector assembly
made from preproduction glass body parts, ALD-functionalized glass
capillary MCP's, an internal resistive High Voltage (HV) divider
implemented with ALD-coated spacers~\cite{HV_paper}, and a microstrip
anode was operated with the laser. The test setup (called the
`Demountable') differed from a true
\LAPPDTM  in that the top-window seal was made with an O-ring
rather than an indium solder seal, the photocathode was a thin film of
aluminum rather than a bialkali film, and the tube was actively pumped
rather than hermetically sealed.  Figure~\ref{fig:APS_Demountable} shows
the Demountable in the APS lab. The blue cables are the readout on the
ends of the anode microstrips and go to 60 channels (30 each end) of
PSEC4 ASICs which digitize the signals at 10
GS/sec~\cite{PSEC4_paper}. The data are then read out via the
FPGA-based ACDC and Central Card PSEC4 DAQ system~\cite{OTPC_paper}.
The Demountable test setup, shown in
Figure~\ref{fig:APS_Demountable}, represents one of three
sub-assemblies of the Supermodule (Figure~\ref{fig:supermodule}).

\clearpage
\newpage
\subsection{Performance}

\begin{figure}[b] 
\centering
\includegraphics[width=0.65\textwidth]{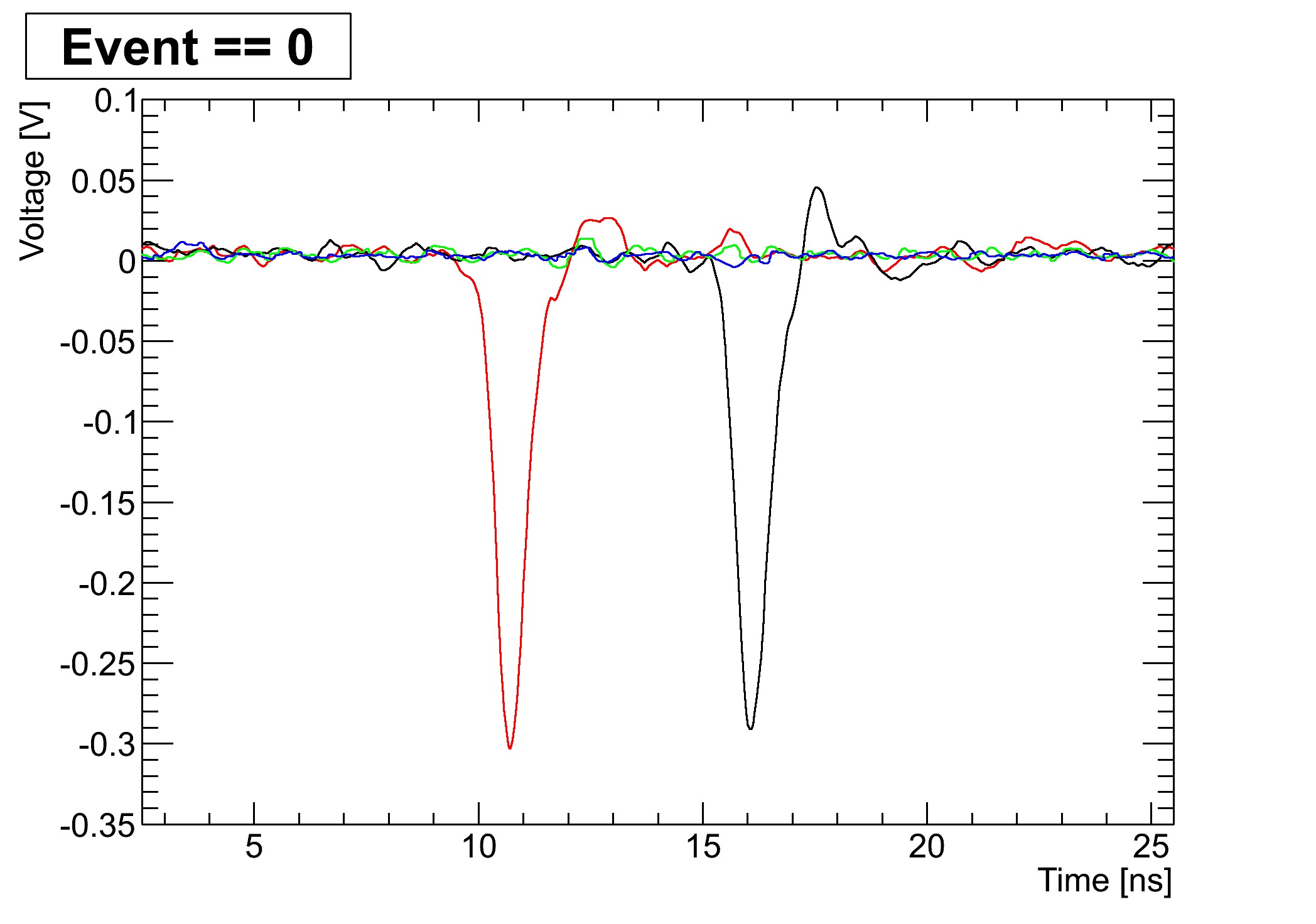}
\caption{The detector response to a single photon, digitized by
the PSEC4 waveform sampling ASIC system, from a complete LAPPD stackup
of metal photocathode, ALD-coated microchannel plates, resistive
spacers, and LAPPD module base with microstrip anode, assembled in the
Demountable test stand at the ANL Advanced Photon Source femto-second
laser lab. The red trace is the pulse measured on one end of the
microstrip transmission line under the laser spot position; the black
trace is the same pulse measured on the other end. The other two
traces are neighboring striplines not excited by the
laser.~\cite{RSI_paper,timing_paper}.}
\label{fig:erico_pulses}
\end{figure}

Figure~\ref{fig:erico_pulses} shows the response to a single
photon from an LAPPD stackup consisting of a metal photocathode, ALD-coated
microchannel plates, resistive spacers, and LAPPD module base with
microstrip anode, assembled in the Demountable test stand at the
ANL Advanced Photon Source femto-second laser lab~\footnote{The
Demountable is a complete LAPPD glass module, but with an O-ring top
seal and metal photocathode.}~\cite{RSI_paper}.  Here the electron
cascade has formed a pulse that propagates in both directions on
the microstrip anode away from the initial charge deposition. Both
ends of each of the 30 strips are digitized using 60 channels of the PSEC4
custom waveform sampling ASIC system~\cite{PSEC4_paper}. The
difference in arrival times at the two ends gives the position
with sub-mm precision; the average of the times gives the time of
arrival.

\begin{figure}[btp] 
\centering
\includegraphics[width=0.75\textwidth]{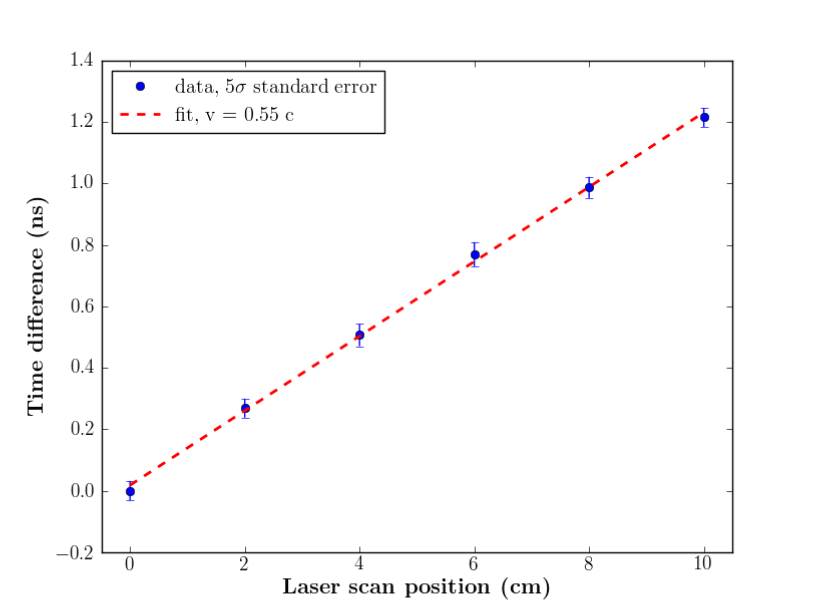}
\caption{The time difference measured in the APS laser lab
between the two ends of the module striplines versus the position of the
laser spot projected on the module~\cite{timing_paper}. Because the
position in the strip direction is calculated from the difference in
times measured at the strip ends~\cite{anode_paper}, 
one expects a linear behavior, as observed. The position in the
orthogonal direction is measured from the charge sharing on
neighboring strips. The resolution in each direction is $\sim$ 700 micron~\cite{timing_paper}.}

\label{fig:time_vs_position_laser}
\end{figure}

Figure~\ref{fig:time_vs_position_laser} shows the time difference
measured in the APS laser lab between the two ends of the module
striplines versus the position of the laser spot projected on the
module. The relationship is linear as expected; the measured resolutions
in both transverse directions are $\sim 700$ microns~\cite{timing_paper}.

\begin{figure}[tbp] 
\centering
\includegraphics[width=0.70\textwidth]{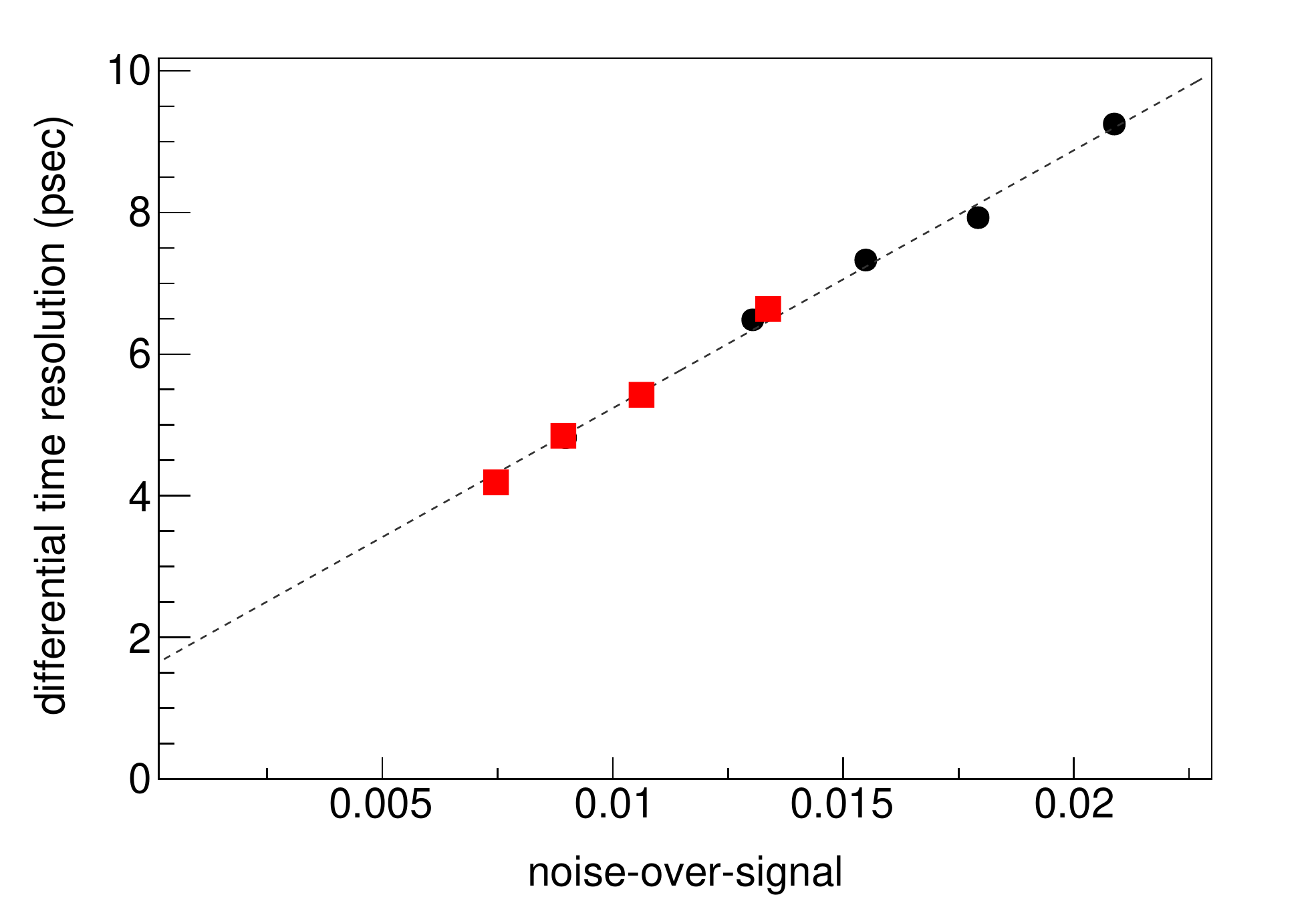}
\caption{The time difference measured in the APS laser lab between the two ends
of the module striplines versus the inverse of Signal-to-Noise in the
8" test chamber in 2012 (black circles)~\cite{timing_paper}.  The
error bars are smaller than the symbols. The signals are large due to
the high gain; the resolution is dominated by noise from the
femto-second laser. The red squares show data taken in 2014 with the
Demountable chamber. The intercept of the fitted line intersects at 1.6
psec. With smaller capillary pores~\cite{Incom_Chris_SBIR} and a signal-to-noise ratio
dominated by the intrinsic noise of the PSEC4 chip we would expect
sub-psec resolution~\cite{PSEC4_paper,timing_paper,Ritt_workshop}. }
\label{fig:money_plot_both}
\end{figure}

Figure~\ref{fig:money_plot_both} shows the time difference
measured in the APS laser lab between the two ends of the module
striplines versus the inverse of the signal-to-noise ratio. Noise at the APS
laser lab was dominated by pickup from the laser pulse generation
itself; the typical noise level of PSEC4 is 700 micro-volts,
negligible in this environment. Later measurements decreased the
intercept, which corresponds to large pulses such as would be
generated by a particle generating Cherenkov light in the detector
window, to $\sim 1$ psec~\cite{timing_paper}.

%
%

\section{Goals Achieved and Not Achieved}
\label{sec:goals_assessment} The LAPPD Collaboration ended in Dec. 2012,
and the R\&D effort largely moved to technology transfer for
commercialization~\cite{Incom_TTO}, as described in
Section~\ref{sec:end_of_project}). Here we briefly summarize the major
goals achieved, the opportunities we identified as viable but were
unable to explore, and the failure to produce a sealed \LAPPDTM module
during the R\&D.

\subsection{Major Goals Achieved}
Major goals achieved include:
\begin{enumerate}
\setlength{\itemsep}{-0.03in} 

\item Establishment of the production capability for high-quality 
8''-square capillary
glass substrates, including fusing into large blocks, slicing and
finishing the blocks into uniform smooth wafers, post-manufacturing
cleaning, and quality control, documentation, and handling
techniques;

\item Development of ALD reactor configurations and
processes to produce adequately uniform ALD coatings for both the
resistive and SEY layers;

\item Demonstration of deposition of 8''-square
 photocathodes on B33 glass substrates for both \KtwoNaSb and
 \KtwoCsSb alkali photocathodes.

\item Development of large-area test facilities that have
been carried over into commercial production facilities; 

\item Extensive
measurements of the performance of 8''-square ALD-coated MCP pairs,
modules, including uniformity, gain, time and space resolution, and
lifetime.

\item Development of the low-cost low-power
6-channel CMOS 15 GS/sec PSEC-4 ASIC; 

\item
Integration and operation with an \LAPPDTM detector of the
PSEC-4-based  60-channel waveform sampling digitizing system with
10 GS/sec at 10.5 bits and 1.6 GHz bandwidth from cathode to DAQ;

\item Development of an
FPGA-based scalable front-end and DAQ system of 30-channel (5
PSEC-4 ASICs) PC cards and a master control card capable of
several psec resolution; 

\item Operation of both the 
pre-production ceramic and glass \LAPPDTM module packages; the
ceramic with full photocathode- the glass with a metal cathode.
The ceramic module was complete but operated in UHV; the glass module
was O-ring sealed and externally pumped~\footnote{The ceramic module tests
were done not long after the end of the Collaboration.}. 

\item Demonstration of \LAPPDTM performance: gain $>10^7$, time resolution of $<50$ psec for single
photons, spatial resolutions of $\sim 700$ microns.

\item Coordination of and securing funding for 
the transition to commercial \LAPPDTM production.
\end{enumerate}

\subsection{Opportunities Missed: What We Did Not Get Done}
\subsubsection{A Sealed Functioning \LAPPDTM Module}
\label{sec:sealed_tile}
LAPPD was not able to successfully fabricate a leak-free sealed
\LAPPDTM module with a bialkali photocathode in the R\&D period, which
ended in December 2012. One trial of making a complete tube with
photocathode was made shortly after the end of the Collaboration. That
first attempt (July 2013) to seal a complete ceramic tile failed when
a spot-weld gave way on one corner of the tile, creating a
leak. However extensive testing done with the tile still in the UHV
tank showed that the tube was fully functional with resolutions in
time and space, gain, and noise rates at or beyond specifications.

\begin{figure}[!h]
\centering
\includegraphics[angle=0,width=1.0\textwidth]{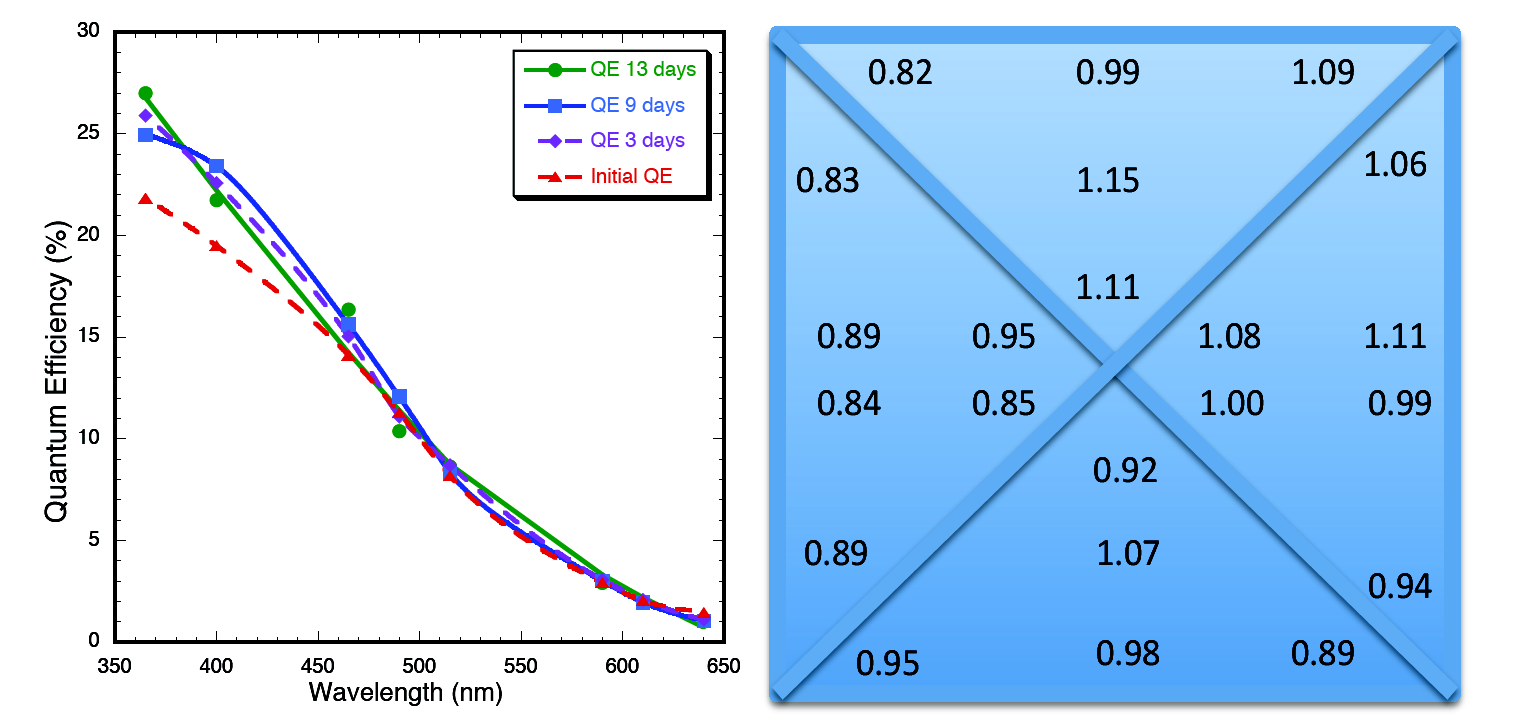}
\caption{Left: Quantum efficiency as a function of wavelength for the
\twentysquare  (Na$_2$KSb) photocathode on the first fully-assembled \LAPPDTM module while
still under vacuum in the SSL tank.  The tube was fully functional,
with good gain and uniformity. When brought up to air one corner of
the seal between the module assembly and the window leaked due to
misalignment. }
\label{fig:ossy_Tube1_QEandUniv_crop}
\end{figure}

Figure~\ref{fig:ossy_Tube1_QEandUniv_crop} shows the characterization
of the 20 cm (Na$_2$KSb) photocathode on the first fully-assembled
\LAPPDTM module
while still under vacuum in the SSL tank (see
Figure~\ref{fig:ossy_big_tank}).  The left-hand panel shows the quantum
efficiency as a function of wavelength. The right-hand panel
demonstrates the uniformity of the photocathode response over the 20
cm by 20 cm area.

\begin{figure}[!h]
\centering
\includegraphics[angle=0,width=0.70\textwidth]{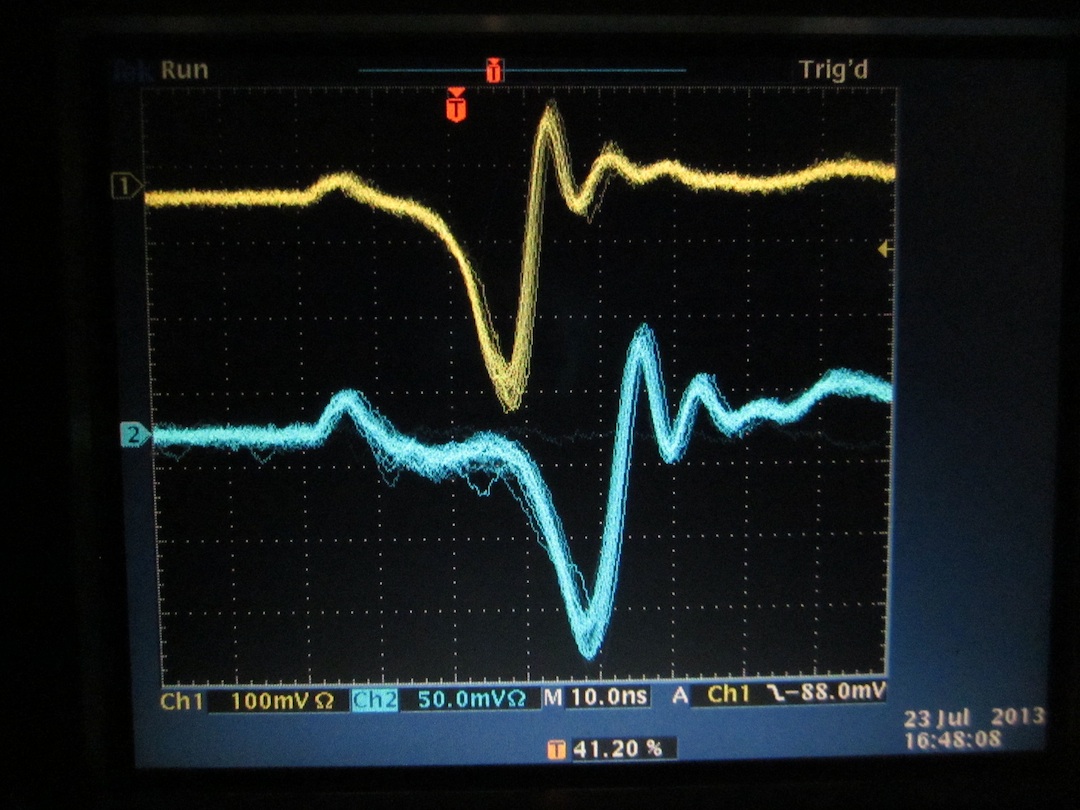}
\caption{The pulses on the anode delay line from the first trial of a 
fully-assembled module while still under vacuum in the SSL tank,
with the photocathode illuminated by a pulsed 610nm laser. }
\label{fig:ossy_Tube1_pulses}
\end{figure}

Figure~\ref{fig:ossy_Tube1_pulses} shows the output pulses from the
delay-line anode of the fully-assembled tube illuminated by a pulsed laser 
while under vacuum in the SSL tank. The bandwidth is limited by the
UHV-compatible electrical connections to the anode; otherwise the
performance of the tube in photocathode response, gain, and spatial
mapping was as expected.

We believe that it is important to understand the sealing trial
history, as it can easily be mis-construed as a fundamental problem with the
design or the sealing technique. Instead, we believe that scaling a
highly-tuned commercial recipe, such as a low-temperature indium seal of a
window with a bialkali photocathode, by a factor of four from the
Planacon dimensions to the \LAPPDTM dimensions will always require
many trials. The lessons learned have now been transferred from SSL to the
commercialization effort at Incom.

Fully integrated trials had been scheduled to be earlier, but were delayed by
unforeseen difficulties, overcome, but with a cost in time. Particular
problems that were solved by the R\&D, but turned out to be
appreciably more difficult than anticipated at the time of the
proposal were: a) the cleaning and handling of the capillary
substrates as the processing techniques were being developed across
multiple geographically-separated institutions; b) the consistent
synthesis of uniform SEY and resistive layers on the 6.5 m$^2$
surface-area capillary wafer,
particularly in an ALD facility shared with other users using
different chemistries; c) subtle packaging problems related to thermal
differential expansion over the much larger dimensions during vacuum
bakeout; and d) slowness/inability by the leadership in re-allocating
resources for the transition from R\&D to production in a wide-spread
diverse organization, well-suited for the required individual R\&D
efforts, but not easily transitioned in budget or expertise onto one
primary task.

\subsubsection{ALD-enabled High Performance Advanced Designs}

The LAPPD R\&D effort concentrated on the basic issues of finding at
least one solution to each problem. In such a process, if successful,
one finds that there are better solutions as well as new opportunities
that were not previously realistically within reach. We can imagine a
future program dedicated to following up on these, or, perhaps more
realistically, a steady chipping away by interested parties (academic,
national labs, industry) with funding from a wide variety of
sources. In particular, the exploitation of the ALD process to use
higher-SEY materials at the pore entrance and discrete dynode
structures~\cite{jeff_rings_patent}; and pushing the limit on the
measurements of timing resolution below 1
psec~\cite{limitations_workshop} remain promising areas in this
technology.

\subsection{Spin-offs from LAPPD R\&D}

There were other opportunities that
spun off parallel R\&D efforts involving individual institutions from
the LAPPD Collaboration, supported by the DOE SBIR/STTR
program or, in the case of medical imaging, a private foundation:
\begin{enumerate}
\setlength{\itemsep}{-0.03in}
\item High quantum efficiency theory-based 
bialkali photocathode synthesis~\cite{RMD_SBIR, Cornell};
\item High bandwidth microstrip and pixel anodes~\cite{Innosys_SBIR};
\item Large-area ALD-functionalized 
MCPs with smaller capillary pores for better time resolution~\cite{Incom_Chris_SBIR};
\item Technology transfer for commercial production~\cite{Incom_TTO};
\item Uniform high-gain ALD coatings~\cite{Arradiance_SBIR};
\item Use of fast timing for reconstruction of particle tracks using
Cherenkov light~\cite{double_beta,OTPC_paper};
\item Use of fast timing for reconstruction of gamma-ray interactions
in water-based PET detectors~\cite{Driskill}.
\end{enumerate}


%
%
%
\section{Major Constructed Facilities}
\label{sec:facilities}

\begin{figure}[b!] 
\centering
\includegraphics[width=.50\textwidth]{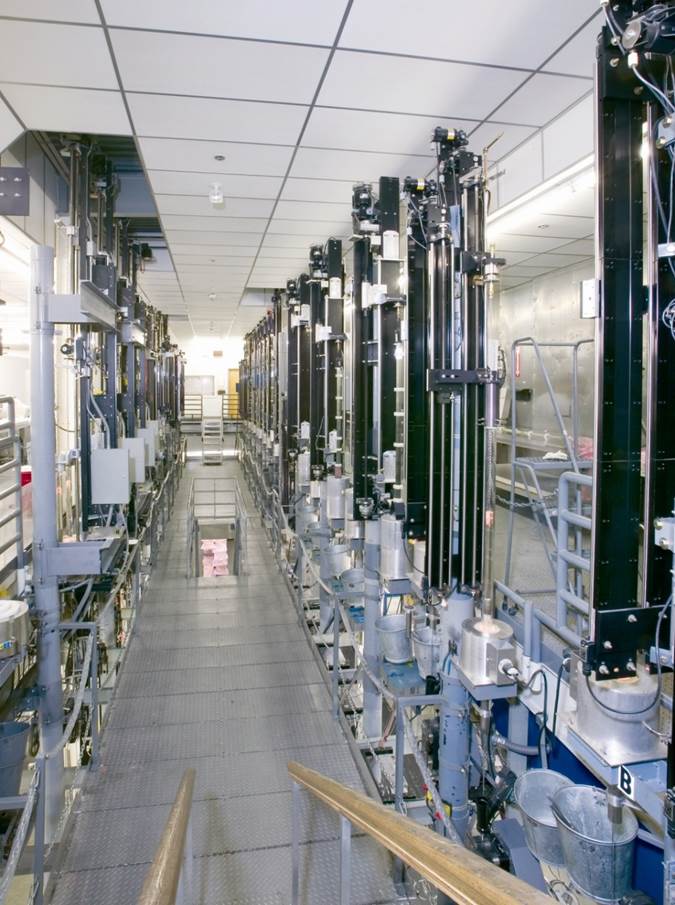}
\caption{The capillary drawing facilities at Incom. The capillaries
 are cut, stacked and fused into a solid block, which is then sliced
 into wafers that are ground, polished, and trimmed to become the MCP
 substrates.}
\label{fig:Incom}
\end{figure}

The prior existence of major facilities at the collaborating
institutions was essential to the project, for reasons of both time,
money, and, most importantly, the availability of expert
personnel. In addition, major facilities dedicated to LAPPD R\&D were
constructed at Argonne, the Space Sciences Laboratory at Berkeley, Incom, and
Fermilab, and diagnostic equipment and software tools were upgraded at
Chicago and Hawaii.  Figure~\ref{fig:Incom} shows the facilities at
Incom~\cite{Incom} for drawing the glass capillary tubes used in the
MCP substrates.  After multiple drawings, fusing into a block, slicing
into wafers on a bias, and finishing, each 20-cm-square substrate
contains $\sim$80-million pores.

\begin{figure}[!ht]
\centering
\includegraphics[angle=0,height=0.25\textheight]{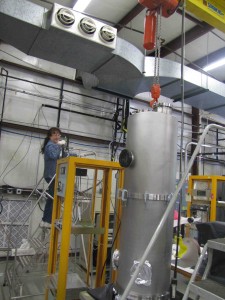}
\hfil
\includegraphics[angle=0,height=0.25\textheight]{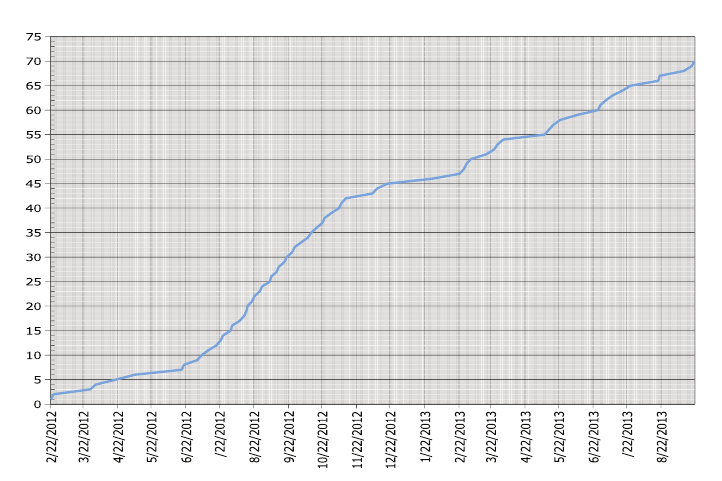}
\caption{Left: The Fermilab UHV evaporation facility 
constructed to deposit NiCr electrodes on the top and bottom surfaces
of \LAPPDTM capillary wafers; Right: A chart of the number of wafers
metalized on both sides at the Fermilab evaporation facility versus time.}
\label{fig:fermilab_chamber_and_flow}
\end{figure}

The top and bottom surfaces of the wafers were metalized with NiCr 
at a custom UHV facility 
constructed at Fermilab to allow uniform evaporation onto 
\twentysquare  plates while rotating them 
about an axis parallel to the pores for proper
end-spoiling~\cite{Wiza_endspoiling}.  The left-hand panel of
Figure~\ref{fig:fermilab_chamber_and_flow} shows the UHV evaporation
facility constructed to deposit NiCr electrodes on the top and bottom
surfaces of \LAPPDTM capillary wafers; In the right-hand panel we
reproduce a Fermilab chart of the number of 8''-square micro-channel
wafers metalized on both sides versus time.

\begin{figure}[btp] 
\centering
\includegraphics[width=1.0\textwidth]{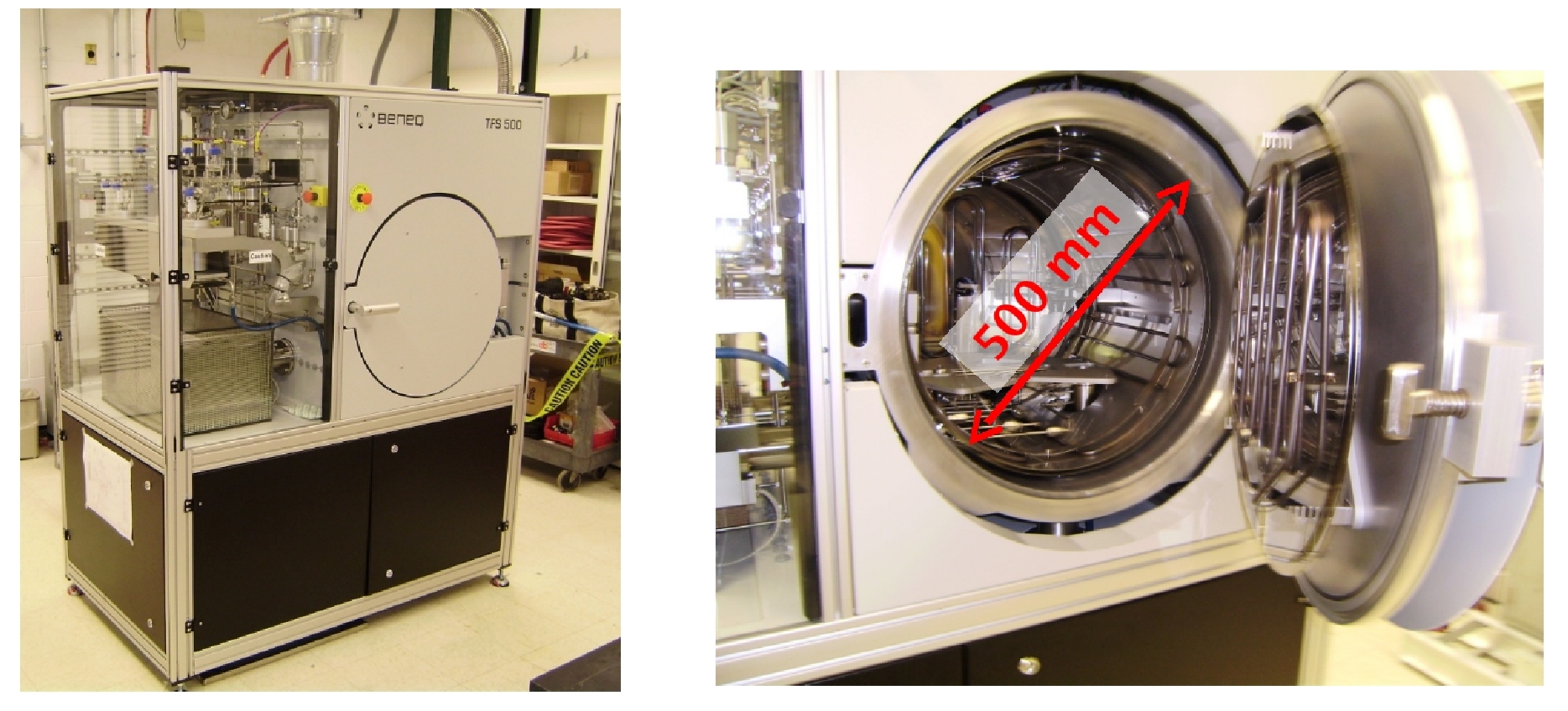}
\caption{The Beneq Atomic Layer Deposition facility installed in
the Energy Systems Division at ANL, used to functionalize the
Incom capillary plates with resistive and secondary-emitting
layers. The precursor flow path required extensive modification in
order to establish uniformity over the 6.5 m$^2$ surface area of
each capillary plate.} \label{fig:Beneq}
\end{figure}

The coating of the glass capillary substrates with resistive and
emissive layers using
Atomic Layer Deposition (ALD) was done in
the Energy Systems Division (ESD) at ANL. Several existing custom
reactors and a large commercial reactor~\cite{Beneq}, acquired at the
start of the project, were shared by LAPPD with other ANL programs for
the development of the secondary-emitting and resistive layers.
Figure~\ref{fig:Beneq} shows the Beneq reactor installed in the ESD labs
at ANL.

\begin{figure}[btp] 
\centering
\includegraphics[width=.75\textwidth]{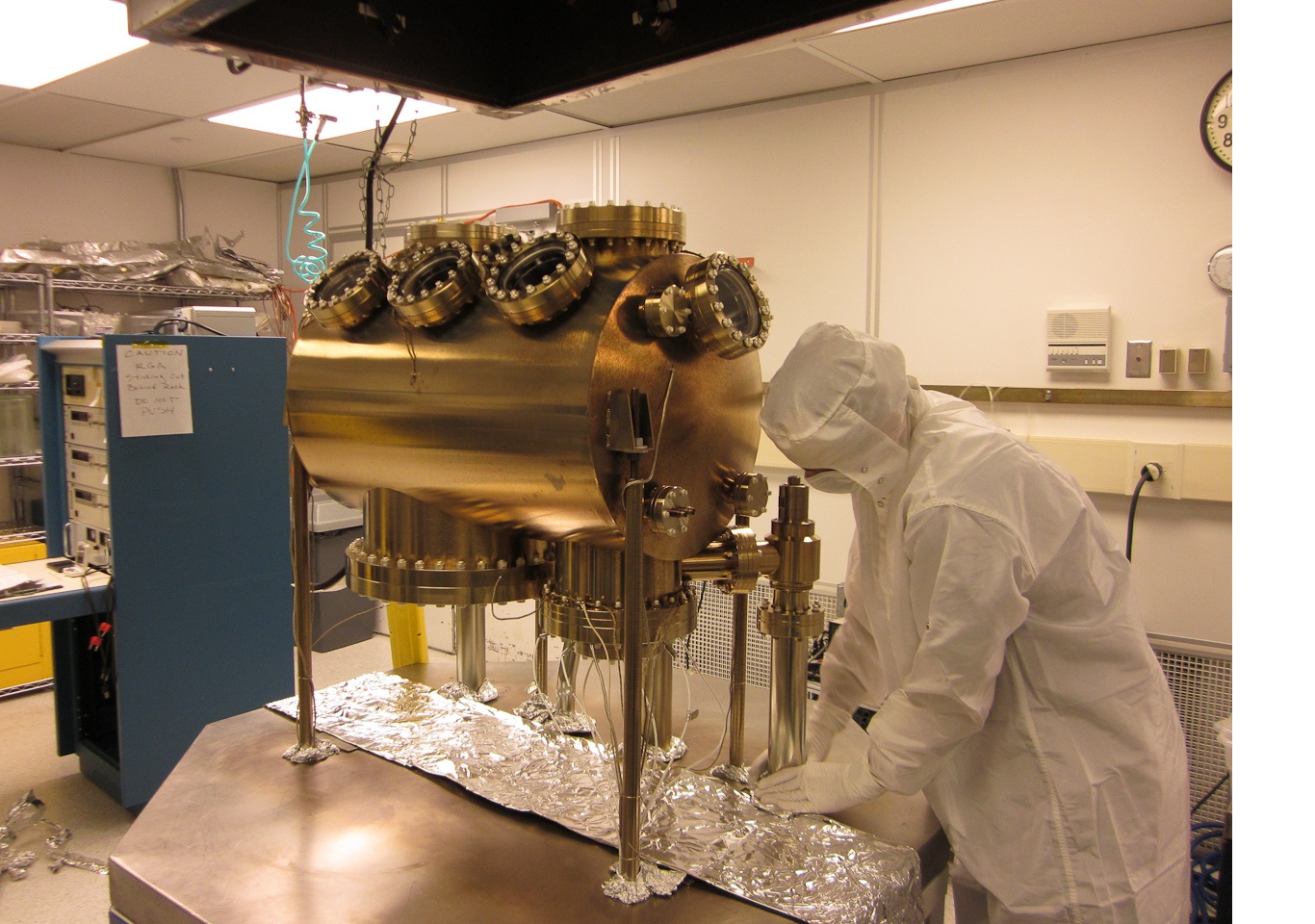}
\caption{The UHV photocathode and detector assembly facility
                    constructed for LAPPD at the \SSLB.}
\label{fig:ossy_big_tank}
\end{figure}

The large process chamber at the Space Sciences Laboratory, shown in
Figure~\ref{fig:ossy_big_tank}, was designed, procured,and commissioned
for the LAPPD program. Large-area photocathode development using a
\KtwoNaSb process established by SSL was done in the photocathode
sub-assembly. The facility is highly instrumented for process control
and subsequent in-situ detector testing.  Modifications were made to
other existing facilities at SSL, including the processing oven,
vacuum baking chamber, the MCP metallizing evaporator, and cleaning
equipment.

The development and characterization of high-quality, reproducible,
glass substrates with uniform high-gain stable ALD coatings required
the building at SSL of extensive testing vacuum facilities to
accommodate the large module format. Testing was initially done with a
standard 33 mm circular format allowing use of fixturing from 
prior SSL programs, and quickly
evolved to the \LAPPDTM \twentysquare plates.  Plates were fabricated
at Incom, sent to Argonne where they were checked and further
documented, and then transported to Fermilab for metalization, and
then transported back to Argonne, where the ALD resistive and emissive
coatings were applied. Plates were then either sent to SSL for tests
of gain, uniformity, and lifetime, or to the test facility at the
Argonne APS for timing measurements, as well as uniformity and
position linearity and resolution with the LAPPD electronics.  At each
step in the `pipeline' an effort was made to perform quality control and to
enter the details in a database that recorded the history of each
microchannel plate.

\begin{figure}[btp] 
\centering
\includegraphics[width=.45\textwidth]{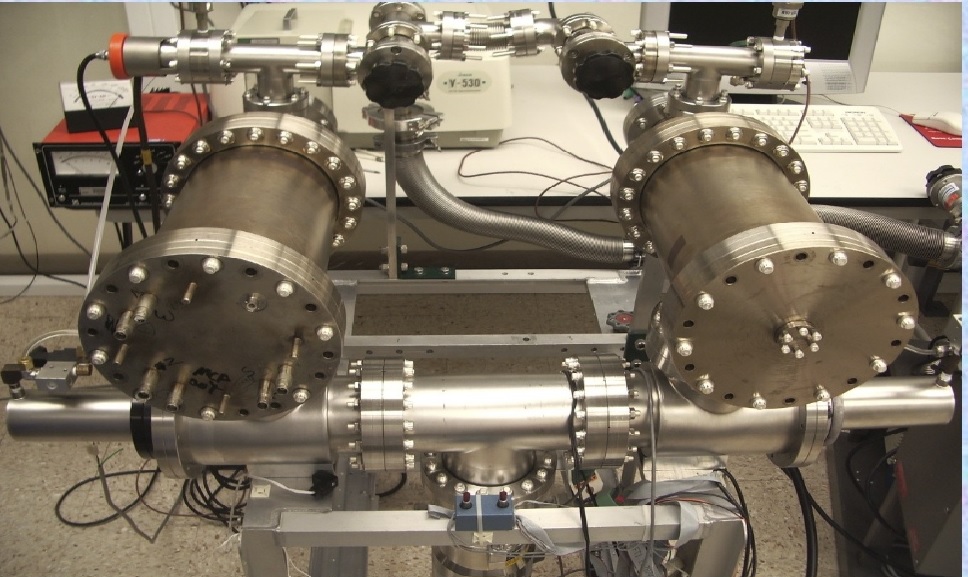}
\hfil
\includegraphics[width=.45\textwidth]{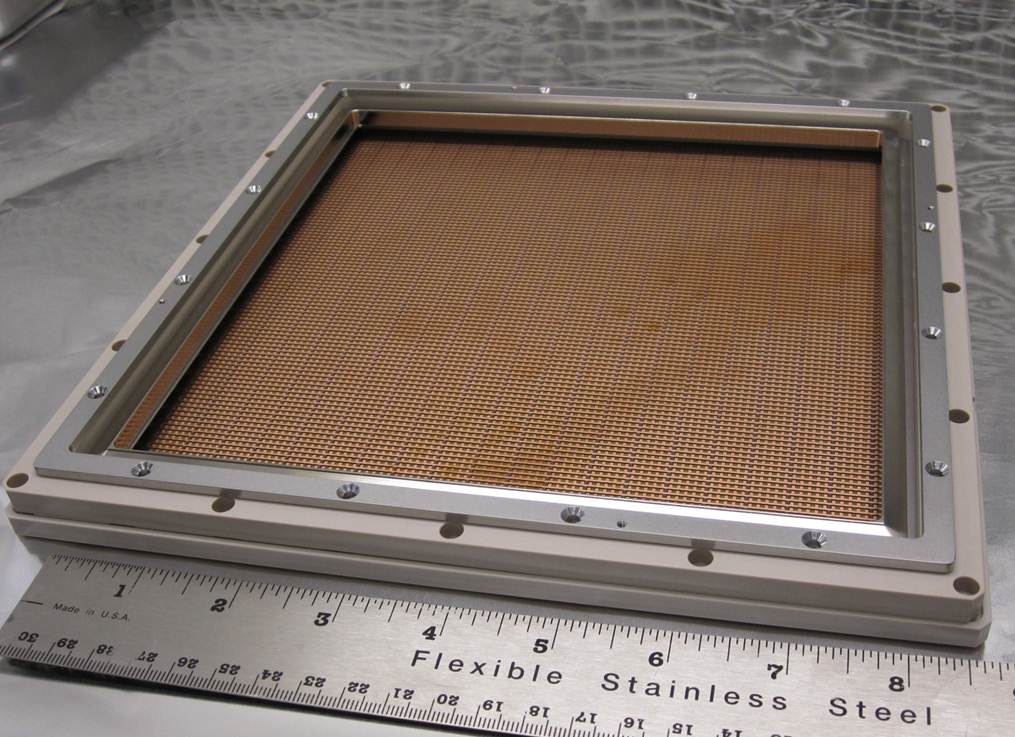}
\caption{Left:: The Dual Chamber Vacuum Test Chamber at \SSLB for rapid
turnaround testing of ALD coatings on 33 mm circular glass substrates
as the processes for both were being developed at the Argonne Energy
Systems Division and Incom, respectively.  This `double barrel test
chamber' employs both phosphor screen detectors for single MCP
characterization and crossed-delay-line detectors for double MCP stack
testing.  Right: The SSL `8''-demountable' vacuum facility designed and built
for characterizing the performance of 8'' ALD-coated MCPs.  The
detector employs a 200mm cross-delay-line anode  for signal readout,
achieving sub-100 $\mu$m spatial resolution, and is designed for easy
MCP replacement to facilitate rapid test cycling.}
\label{fig:ossy_test_setups}
\end{figure}

Figure~\ref{fig:ossy_test_setups} shows two test facilities
constructed at SSL for LAPPD that were essential in the
development of uniform high-gain low-noise ALD coatings on glass
substrates of high uniformity and very few blemishes. The left-hand
panel shows the `Dual Chamber Vacuum Test Chamber' used for rapid
turnaround testing of ALD coatings on the 33 mm circular glass
substrates, of which many could be made from a single 8'' wafer. The
right-hand panel shows the `8''-demountable' detector, consisting of a
high resolution cross-delay-line anode mounted on a large vacuum
vessel, designed for rapid test cycling. This instrument was the
`work-horse' facility for the development of the 8'' Incom glass
substrates and the Argonne ALD coatings, providing measurements of
gain uniformity, resolution, and noise.

\begin{figure}[btp] 
\centering
\includegraphics[width=.45\textwidth]{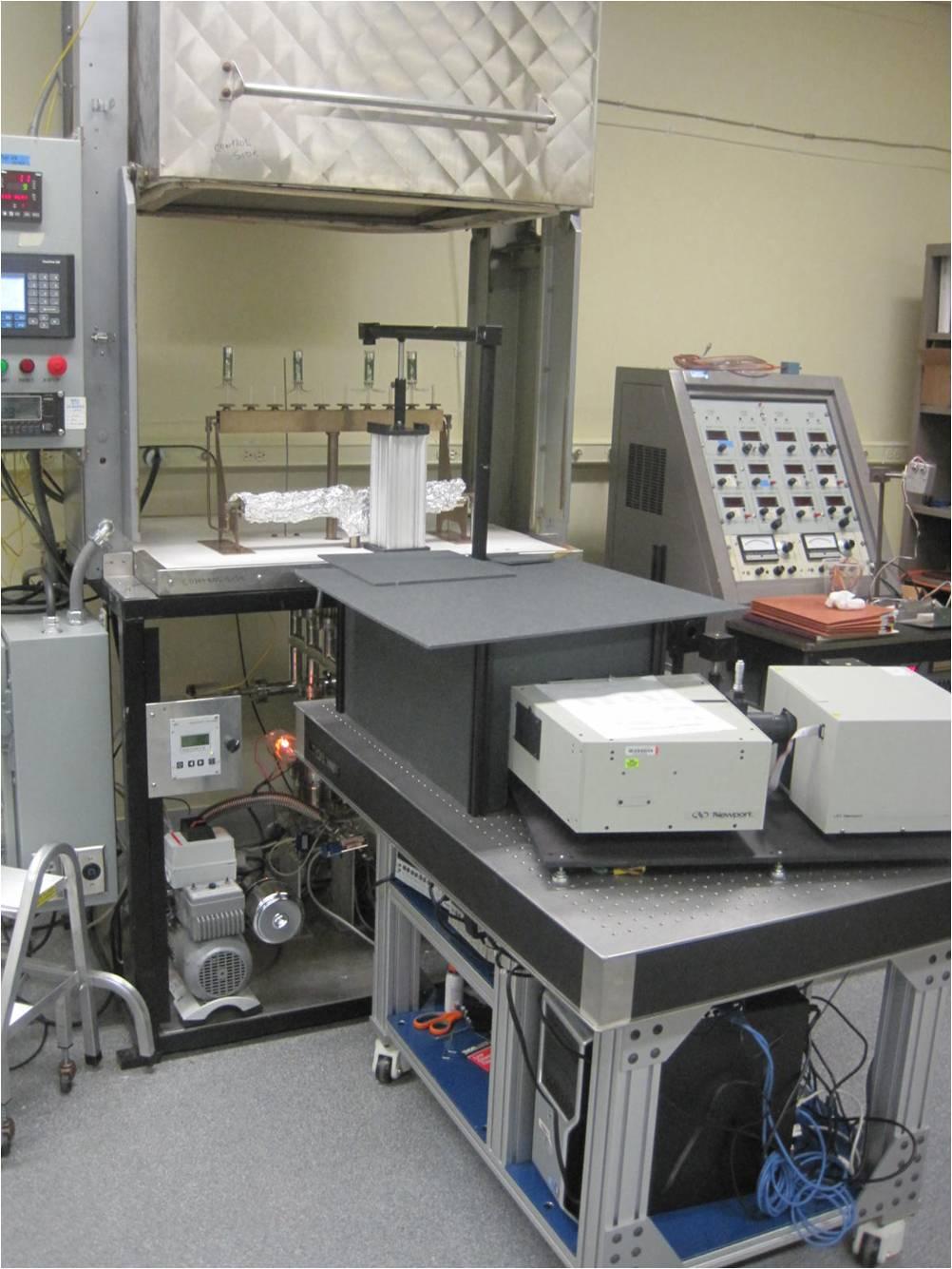}
\caption{The commercial Photocathode Deposition facility, purchased
from Burle/Photonis,
installed in the High Energy Physics Division at ANL. The
instrumentation used for measuring film thickness and quantum
efficiency is in the foreground.} \label{fig:Burle}
\end{figure}

The Chicago group had been
working with commercial Photonis Planacons~\cite{Ossy_Planacon} in
the Fermilab  beam tests, and so it was natural to work on the
extension of the Planacon \KtwoCsSb cathodes to the 16-times
larger area \LAPPDTM format. Figure~\ref{fig:Burle} shows the
Burle/Photonis photocathode deposition facility purchased by LAPPD
and installed in a new lab created in the HEP Division at ANL for
photocathode formation and characterization. In order to make
8''-square cathodes, the manifold that held multiple conventional
PMT's was replaced by a large glass vessel, created by J. Gregar
of the ANL Glass Shop, that contained the Sb beads, alkali
sources, and that held the window on which the cathode was
deposited. Diagnostic and characterization equipment were added to
the installation, as shown in Figure~\ref{fig:Burle}.

\begin{figure}[b!] 
\centering
\includegraphics[width=.65\textwidth]{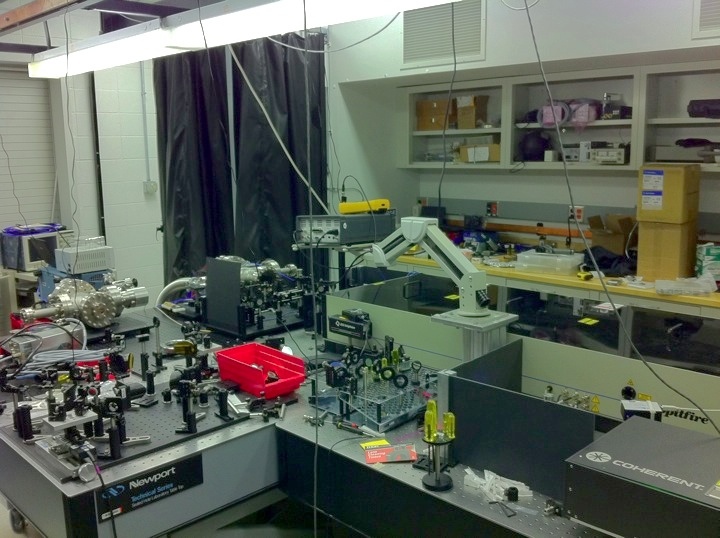}
\caption{The timing lab constructed at the ANL
APS~\cite{RSI_paper} The Ti-Sapphire laser is on the lower right; two
vacuum test chambers are visible on the middle left. The laser beam can
be transported to several test chambers as well as the
`Demountable' prototype module by components on the optical bench.}
\label{fig:APS_lab}
\end{figure}

In addition to the `pipeline' program of MCP development at \SSL
~\cite{Ossy_IEEE_2011,Ossy_JINST_2013}, MCP testing was done at the
Advanced Photon Source~\cite{RSI_paper,timing_paper}, with the
emphasis at SSL being gain, uniformity and lifetime, and with
Chicago/ANL concentrating on fast timing.  Figure~\ref{fig:APS_lab}
shows the femto-second laser laboratory at the APS that was built up
as the test facility for fast timing~\cite{RSI_paper}. A Ti-Sapphire
laser capable of 30 fsec pulses and enough intensity to excite a metal
cathode was refurbished by LAPPD at a modest cost.

\clearpage
\section{The Role of DUSEL in LAPPD}
\label{sec:dusel}
The US joint DOE-NSF proposal for a large underground water Cherenkov
neutrino detector to be installed in the DUSEL laboratory required a
large investment in photodetectors, more than
100M\$~\cite{DUSEL,DUNE}. This served as an impetus for developing
large-area `frugal' planar detectors made of inexpensive glass
components, and given the large payoff for success, was a key
motivation for DOE funding of a high-risk endeavor such as
LAPPD~\cite{Howard_Nicholson}.

\begin{figure}[tbp] 
\centering
\includegraphics[width=0.85\textwidth]{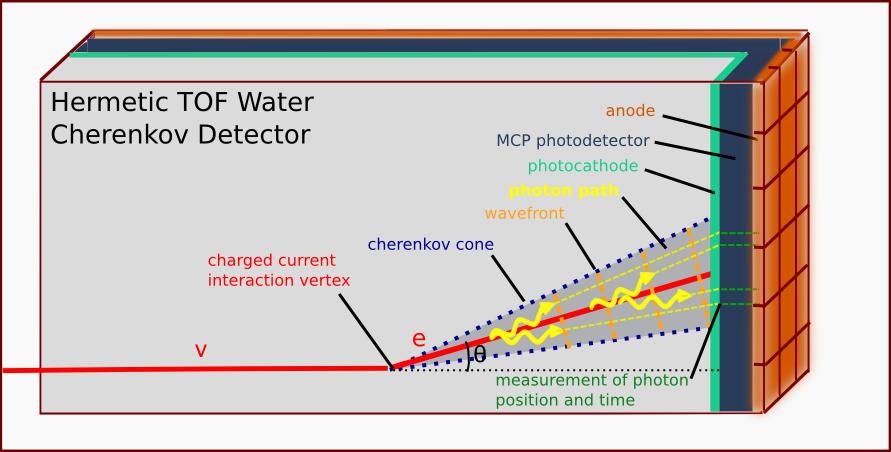}
\caption{A 2008 conceptual drawing of a large water Cherenkov
counter such as was proposed for DUSEL, in this case instrumented
with LAPPDs. Note the use of time and position measurements on
each Cherenkov photon to locate the track~\cite{OTPC_paper}.
(credit: Constantinos Melachrinos) } \label{fig:Constantinos}
\end{figure}

Figure~\ref{fig:Constantinos} shows a 2008 conceptual drawing of a
large water Cherenkov counter such as was proposed for DUSEL, in
this case instrumented with LAPPDs. Note the use of time and
position measurements on each Cherenkov photon that allow using
the photon drift times to reconstruct the track~\cite{OTPC_paper}.

\begin{figure}[tbp] 
\centering
\includegraphics[width=1.0\textwidth]{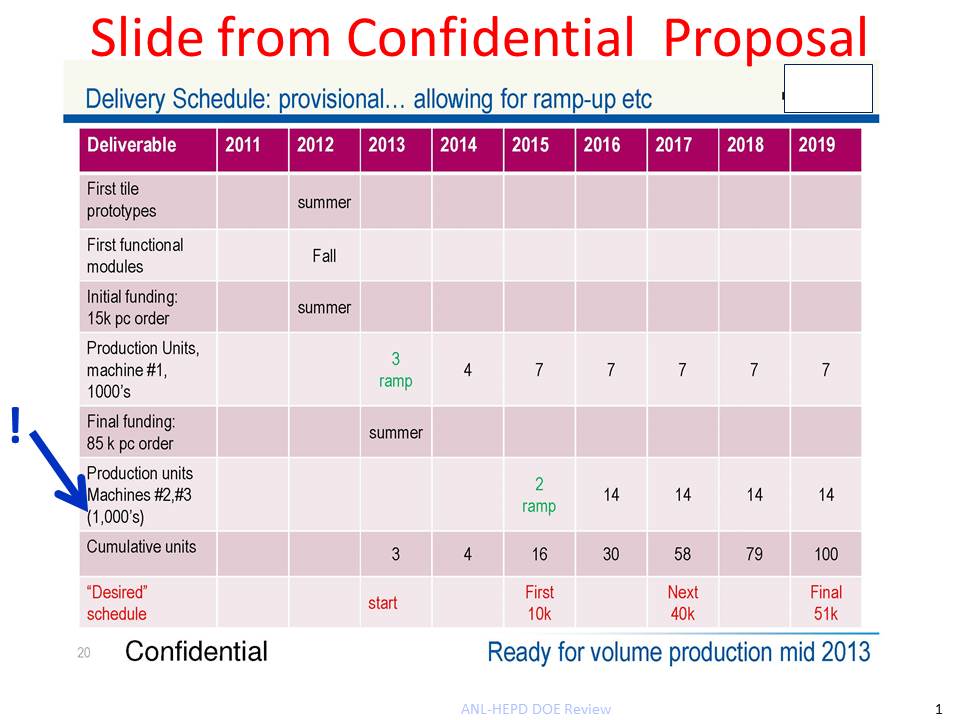}
\caption{A slide from a proposal for production of LAPPDs for
DUSEL~\cite{DUSEL}  by a company experienced in vacuum phototube production. The
proposal was to modify an existing production line and then
replicate it to produce  20,000 LAPPD modules per year. The
commercial interest ended with the cancellation of the DUSEL
project. The arrow is to emphasize that the quantities are in units of
one thousand modules.}

\label{fig:Dusel}
\end{figure}

The large water Cherenkov counter for DUSEL was important in
another aspect as well: the large amount of money involved for
photodetectors attracted the attention of a company that had a
facility that could accommodate the size of the LAPPD tubes, and
produce them in volume. The resulting proposal, from a company
already producing a similar product, provided a strong
confirmation of the goals and methods of LAPPD~\cite{NDA}.
Figure~\ref{fig:Dusel} shows a slide from the proposal.
Unfortunately, the large neutrino detector for DUSEL was canceled
by the US National Science Board~\cite{DUSEL} and the commercial
interest ended~\cite{DUNE}.

\section{Communicating Technical Information in a Multi-Disciplinary Effort}
\label{sec:communicating}
A non-technical but critical goal of LAPPD was to convince the
scientific and commercial communities of the feasibility of large area
timing with resolution measured in psec. This was necessary to secure
funding, to enable commercialization, and to build a fast-timing
community intellectually invested in understanding and developing the
detector and supporting technologies such as low-power psec
electronics digitization
systems, and clock distribution. This communication effort was
time-intensive, with  multiple weekly meetings, international
workshops, interfacing to phototube and other vendors, and extensive
web pages, as described below.

\subsection{Collaboration Meetings} 
\label{collaboration_meetings}
The LAPPD Collaboration was often described as a 'pick-up ball game',
in that the group was self-assembled based on interest, and covered a
very wide range in cultures and methods of working. An unusual amount
of effort was consequently needed to make efficient progress.
LAPPD Collaboration Meetings were held twice-yearly to cover the
broad range of material science R\&D, proprietary knowledge, and
education of those from other fields. The agendas and slides
are available on the LAPPD web page~\cite{collab_meetings}.

\subsection{Designing an Effective Review Process}
In addition to the collaboration meetings, each of the four areas of
R\&D reported to its own review committee, dubbed 'Godparent
Review Panels'. The panels met twice per year, out-of-phase with
the twice-yearly Collaboration meetings. Each panel was made up of both outside
experts and, in an effort to avoid the typical reviewer/reviewee
dynamics,
 LAPPD members who were not participants in the area under review,
bringing to the conversation both outside experience and some
knowledge of why internal choices that may have been different
from conventional wisdom were made. In addition, an effort was
made to bring in critics of the overall concept.

The godparent reports and 
makeup of the initial round of godparent panels are
posted on the psec web Library page~\cite{psec_library}; the
procedure called for a written response by the proponents.

\subsection{Collaborative Cross-Community Workshops}
\label{workshops}
A homogeneous amplification section with a
dimension measured in tens of microns should provide fast
pulses capable of psec-level time
measurements~\cite{Aspen,Credo,Ohshima,Photek}. However there remained
many detailed questions. A series of workshops, held alternately in the
Chicago area and in France, was intended to provide answers to these
specific questions one-at-a-time; we first determined the talks
needed, and then worked to identify the most expert speakers to that
topic, often inviting specialists we did not know. Rather than having
speakers report on their own work, the workshop focused on bringing to
bear their expertise on the questions and problems.  The result was a
broadening of the base of both knowledge and community.

\begin{figure}[tbp] 
\centering
\includegraphics[width=0.90\textwidth]{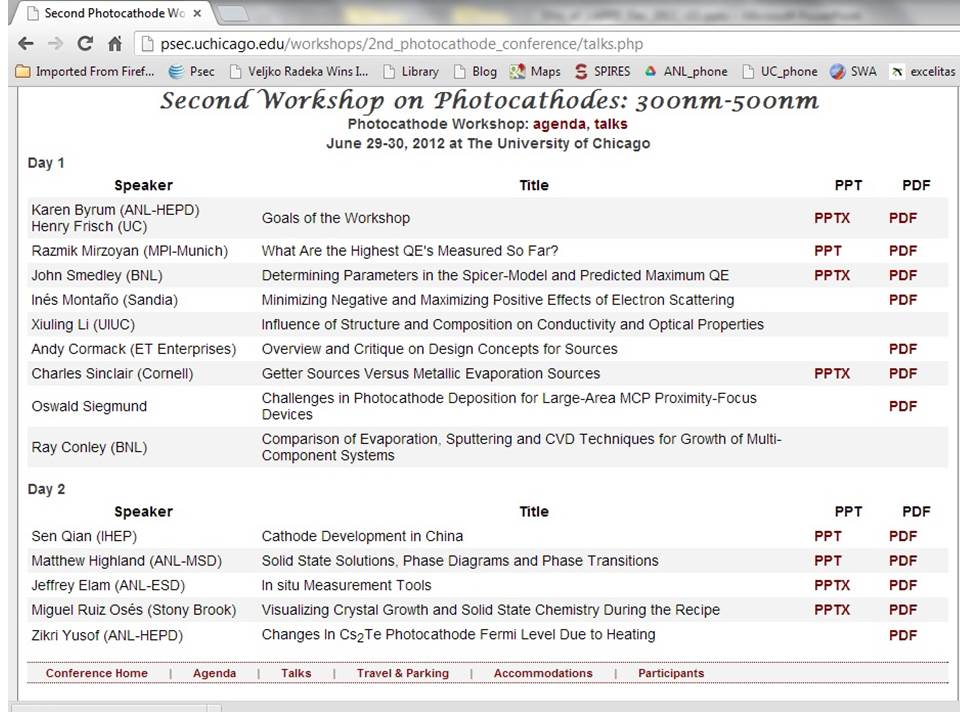}
 \caption{The web page for the {\it Second Workshop on Photocathodes:
 300-500nm}, held at UC in June 2012, one of a bi-annual series held
 in collaboration with P. LeDu and other French colleagues. Note the
 participation from many communities who normally have little
 interaction. The two cathode workshops have been followed up by several
 collaborative efforts~\cite{RMD_SBIR,Cornell}.}
 \label{fig:workshops}
\end{figure}

The agendas and talks at the Workshops are available on the PSEC
Library web page~\cite{workshops}. The workshop on Limitations on
Fast Timing and the two workshops on bialkali photocathodes in the
300-500 nm range are particularly good examples of the
effectiveness of the top-down assignment of titles and
speakers. Figure~\ref{fig:workshops} shows, as an example, the agenda
of the second workshop on photocathodes.

\subsection{Web-based Documentation for Collaborators, Vendors, and Adopters}
\label{sec:web}

 In addition to being the repository for proceedings of the Godparent
Review Committees, the Collaboration Meetings, and the Workshops, the
LAPPD web pages provided the basis of the weekly status reports,
papers and conference proceedings, and often-used information on
materials and techniques.  The Library web
pages~\cite{general_library} also serve as a convenient
easily-accessed source for external material that is often referred to
by detector developers, such as tables of CTE's and moduli of
materials, relevant literature on MCP's, photocathodes, and detector
performance.

\begin{figure}[tbp] 
\centering
\includegraphics[width=0.90\textwidth]{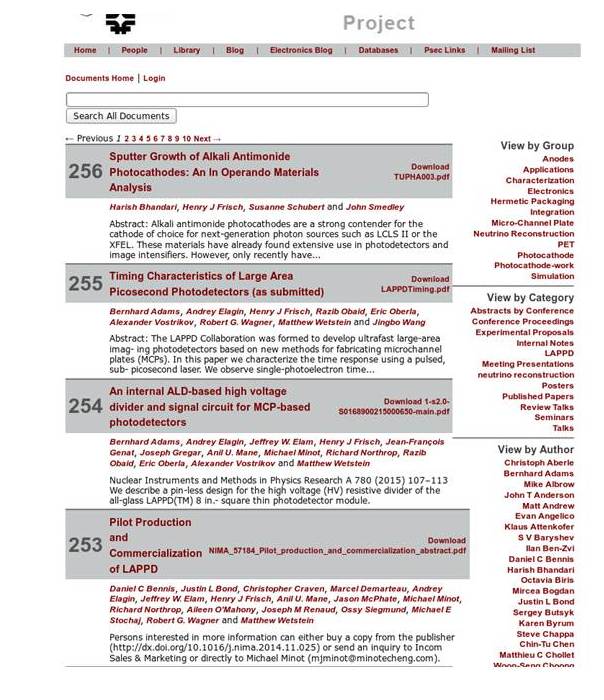}
\caption{A sample page from the Document Library. The data base is
searchable by author, topic, and document type. The Document Library
serves both as an archive of results, and has proved useful in
interactions with potential adopters, other groups working on fast
timing or photodetectors, and prospective students, postdocs, and
employees.  }
\label{fig:doc_lib}
\end{figure}

Three custom web-based tools, open to the public, turned out to be
heavily used for documentation and communication.  The Collaboration
Blog provided a forum and archive for test results, schedules, and
presentations viewed during the weekly phone meetings~\cite{blog}. In
addition, three `libraries' were linked to the web page: the Document
Library for the Collaboration papers, talks, and internal
notes~\cite{doc_lib}; the Image Library for discussing technical
specifications with vendors and archiving prints for
reference~\cite{image_library}; and the Figures Library for use in
talks and papers~\cite{figures_library}. Figure~\ref{fig:doc_lib}
shows a sample page from the Document Library. The Electronics Group,
consisting of members from the University of Chicago and University of
Hawaii, met separately weekly (the meeting was open to all members,
however) and also maintained dedicated web documentation and its own
blog~\cite{electronics_pages}.

\subsection{Areas of Application and Early Adopters: the LAPPD
Strategy}
\label{sec:early_adopters}

\begin{figure}[tbp] 
\centering

\includegraphics[width=0.85\textwidth]{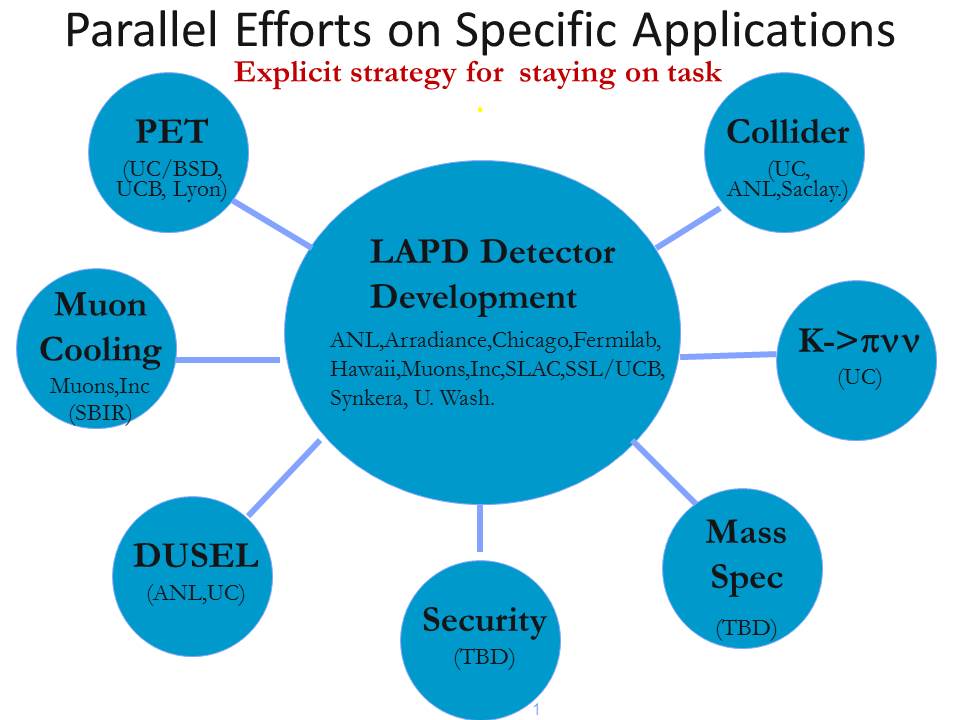}
\caption{A pictorial representation made in 2010, shortly after the start of 
the Collaboration, of the strategy for matching user requirements with 
projected detector capabilities in areas of possible application of LAPPD
photodetectors. The LAPPD core effort was focused on the photodetector
R\&D; specific applications were pursued by individual collaborators
working with outside interested groups. In 2013, after the
Collaboration had ended, a meeting of `Early Adopters' from these
areas was organized by Incom~\cite{Incom_early_adopter_meeting}.}
\label{fig:blue_parallel_efforts}
\end{figure}

The model for LAPPD involvement in applications was to let individual
members follow their interests, often with prior external
associations.  Figure~\ref{fig:blue_parallel_efforts} shows a
pictorial representation from 2010 of this strategy for feedback from,
and information flow to, areas of possible application of LAPPD
photodetectors. In 2013, after the Collaboration had concluded, Incom
organized a meeting of `Early Adopters' from these areas to better
understand specifications and uses~\cite{Incom_early_adopter_meeting}
as part of their commercialization plan. Specific areas that followed
this pattern are the Optical Time Projection
Chamber~\cite{OTPC_paper}, ANNIE~\cite{ANNIE}, Positron-Emission
Tomography~\cite{Heejong_NIM_2009,Heejong_NIM_2010,Heejong_NIM_2011},
and non-proliferation monitoring detectors~\cite{jocher}.

\subsection{Publishing}
While the development of large-area psec photodetectors and
electronics has been the subject of a large program, there are many
areas that are fertile for further development. We encountered many
industrial `recipes' ripe for a deeper understanding of the underlying
material science, chemistry, and physics. LAPPD has consequently put a
high premium on publishing technical details of the development,
including papers on the glass/ALD
MCPs~\cite{AAO_patent,Jeff_ALD_ECS_2013,
Anil_ALD_ECS_2014,ALD_patent,
Valentin_NIM,
Insepov_Ivanov_SEY,
Baryshev_Igor_Jeff_profiling,
Slade_SEY_NIM,Slade_NIMB}, fast timing and
electronics~\cite{RSI_paper,PSEC4_paper,JF_NIM,
anode_paper,Hawaii_electronics}, and
packaging~\cite{Ossy_JINST_2013,HV_paper,anode_paper,Craven_TIPP,Minot_NIM}.
Conference reports from throughout the R\&D can also be found in the
Document Library.

In addition to published papers, patents were
filed on the technical developments in the MCP and packaging areas,
and in uses for HEP/Nuclear Physics and medical
imaging~\cite{1st_patent,
AAO_patent,ALD_patent,Patent_2,PET_patent}.

\subsection{Mentoring, Awards, Careers}
Detector development is both critical to the future of hard
sciences and medical care, and yet is widely considered a
second-tier and hence dangerous career path for young scientists.
This paradox is often discussed, but funding pressures,
particularly on universities from which the young scientists come,
severely constrain any change. One of the products of the LAPPD
project was the recognition of several young scientists: LAPPD
postdoctoral fellows were awarded the Lee Grodzins Award
(Wetstein), the Charles Townes Fellowship (Ertley), and the
Grainger Postdoctoral Fellowship (Wetstein); an LAPPD graduate
student (Oberla) won the Grainger Graduate Fellowship; and a high
school student working with LAPPD took second place in the
national Intel Science Competition (Credo). Senior members were
awarded an R\&D 100 award. In addition, a large number of REU and
SULI students took part in hands-on research in the summer, with
several returning as graduate students or postdocs working on
LAPPD.

%
%
\section{End of Collaboration: Path to Commercialization and Adoption}
\label{sec:end_of_project} 
\begin{figure}[tbp] 
\centering
\includegraphics[width=1.0\textwidth]{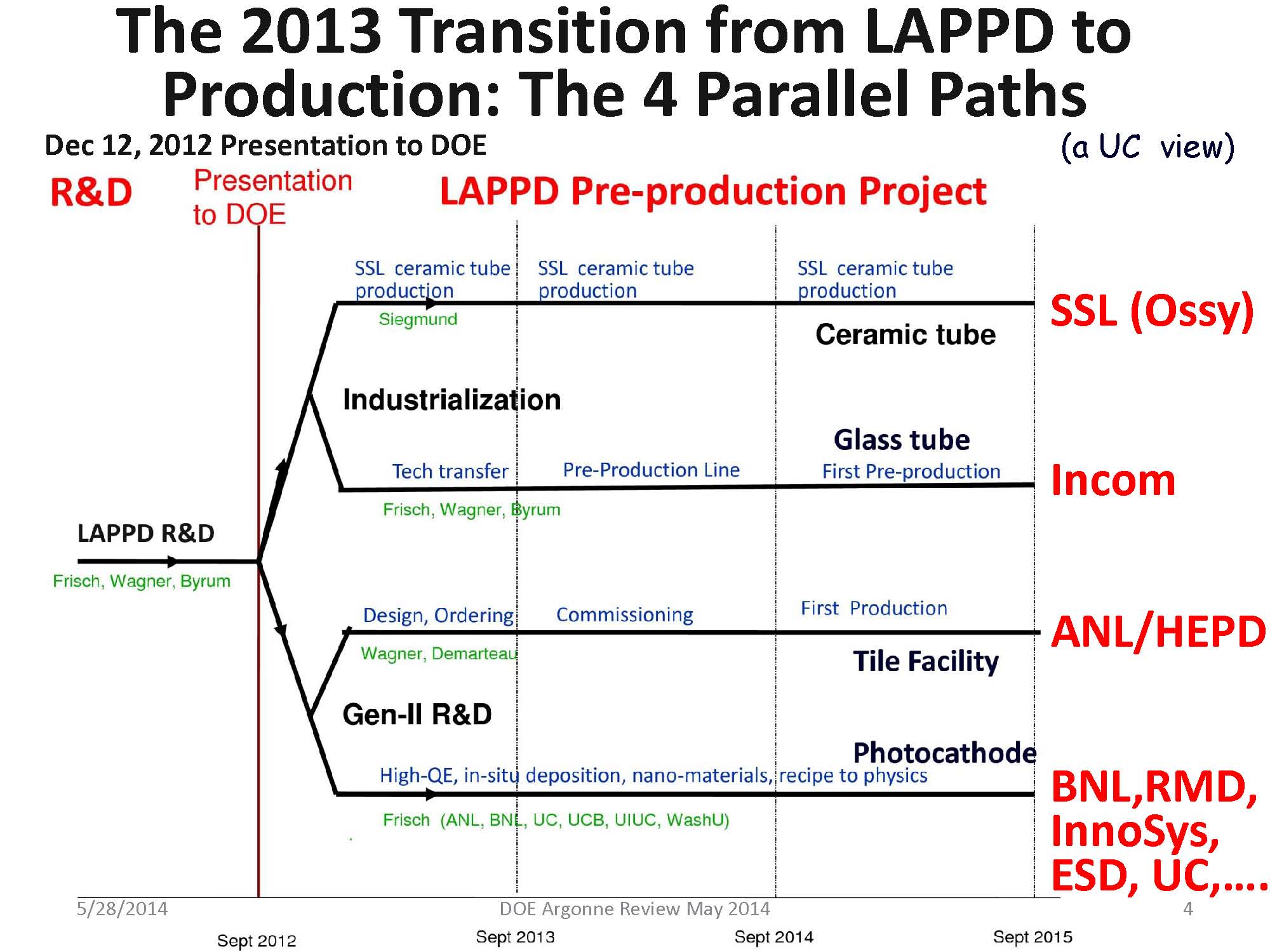}
\caption{Slide from the DOE review at ANL on Dec. 12, 2012,
showing the end of the collaborative R\&D and the transition to
four parallel, separately funded, and individually managed
efforts.}
\label{fig:harry_4path}
\end{figure}

The conclusion of the ARRA funding in 2012
ended the Collaboration structure of university support through
subcontracts from ANL and the organization structure shown in
Figure~\ref{fig:org_chart}. Figure~\ref{fig:harry_4path}, from the
final review of the Collaboration in Dec. 2012, shows the transition
to four parallel, separately funded, and individually managed paths.
The purpose of two of the paths was the transition to
commercialization: Incom~\cite{Incom}, under a DOE Technology Transfer
Opportunity award~\cite{Incom_TTO}, has moved forward on commercial
production of both the MCP's and also the LAPPD$^{TM}$
detectors~\cite{LAPPDTM}, and the SSL group continued on the ceramic
package program under separate funding. The goals of the other two
paths, the subject of collaborative efforts with industry, were to
continue R\&D on technical avenues that had been identified in the
LAPPD project: the bulk synthesis of pure bialkali photocathode
material for sputtering onto substrates~\cite{RMD_SBIR}, higher
bandwidth glass capillary substrates~\cite{Incom_Chris_SBIR},
higher-bandwidth anodes~\cite{Innosys_SBIR}, and developing techniques
for bringing the yield up and the price down through adopting
techniques for conventional PMT production~\cite{Patent_2}.

Lastly, the applications of psec timing have become much better
understood, with growing interest in many scientific and commercial
applications. The use of the measurement of the coordinates of
arrival of photons or charged particles in both time and space has
been explored in water Cherenkov counters~\cite{OTPC_paper}, medical
imaging~\cite{Heejong_NIM_2009,Heejong_NIM_2010,Heejong_NIM_2011,PET_patent},
searches for neutrino-less double-beta decay~\cite{double_beta},
and non-proliferation reactor monitoring~\cite{jocher}. At Fermilab
the ANNIE experiment~\cite{ANNIE} is being constructed to make first
tests of the use of LAPPD$^{TM}$ modules, much as was envisaged for DUSEL.

%
%
\acknowledgments
\label{sec:acknowledgements}
We thank Glen Crawford, Division Director of the Office of High Energy
Physics of the DOE for his essential support, Howard Nicholson and
Peter Kim for their contributions as involved and knowledgeable DOE
program officers.  The \UofCsp group is grateful for funding from the
Driskill and Grainger Foundations and support for \UofCsp personnel
by the National Science Foundation. Robert Fefferman, Dean of the
\UofCsp Physical Sciences Division, provided the critical seed money
at the start of the project.  We are also deeply grateful to Michael
Detarando, President of Incom, who provided crucial support and
resources.

We thank Alfred Sattelberger (Associate Laboratory Director, Physical
Sciences and Engineering Directorate) for a crucial rescue of the ANL
Glass Shop, and to Harry Weerts (Director, High Energy Physics
Division) and Marcel Demarteau (Detector Group Leader, High Energy
Physics Division) for their support during this phase of the
project. We thank J. V'avra for his many intellectual contributions to
the project in the early stages. We are also grateful to Michael
Pellin, who provided essential wisdom and knowledge on material
science throughout the project, and Eric Delanges and Stefan Ritt who
volunteered their expertise on fast sampling electronics. We thank
Dmitri Routkevich for his contributions to the AAO substrate effort at
the start of the project. We are grateful to Jeffrey Defazio and Emil
Schyns of Photonis for their interest and help, and to our excellent
vendors who supported the ceramic and glass packaging efforts.
Lastly, we deeply thank the many staff members at our home
institutions who made the overall effort possible.

\section*{Appendix A}
\begin{figure}[p!] 
\centering
\includegraphics[width=1.0\textwidth]{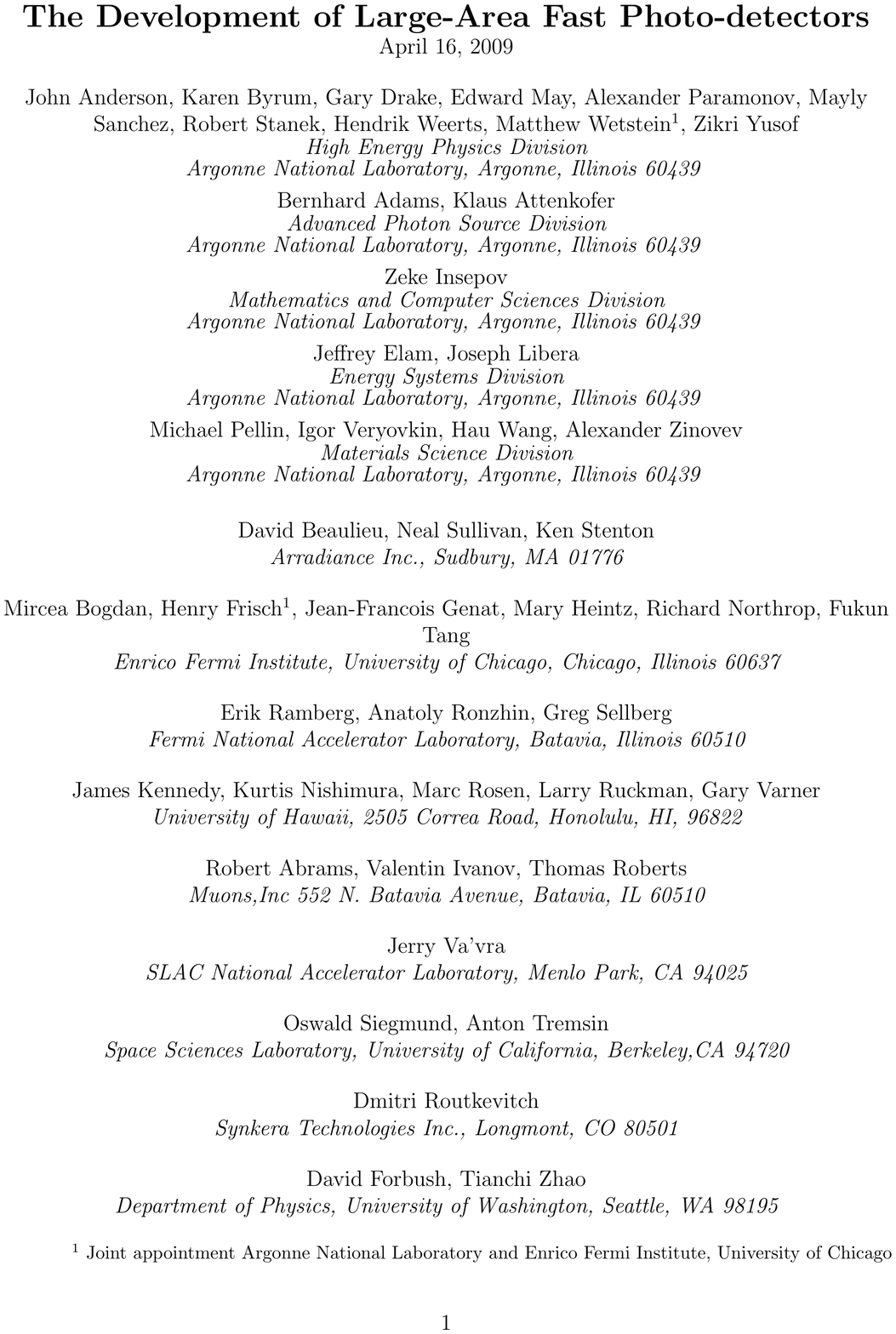}
\caption{The original set of LAPPD institutions and authors on the
2009 Proposal to the DOE. The Collaboration was self-assembled from three
kinds of institutions: National Laboratories, US-based industry, and
universities. Because the proposal was self-generated by the group and not
programmatically driven, it was based on a
`bottoms-up' intellectual interest in specific
aspects in their own areas of expertise. } 
\label{fig:author_list}
\end{figure}

Figure~\ref{fig:author_list} is the page from the 2009 proposal
submission to the DOE Office of High Energy Physics listing authors
and institutions. The LAPPD Collaboration was unusual in that it was
formed with participants from three complementary kinds of
institution, national laboratories, universities, and US companies as
equal partners. In addition, the self-generated inception and the
strong support from the Office of High Energy Physics of the DOE
attracted expertise in many unanticipated scientific or technological
areas, talent that might not have become involved in a more
conventional collaboration.

\begin{figure}[b!] 
\centering
\includegraphics[width=1.0\textwidth]{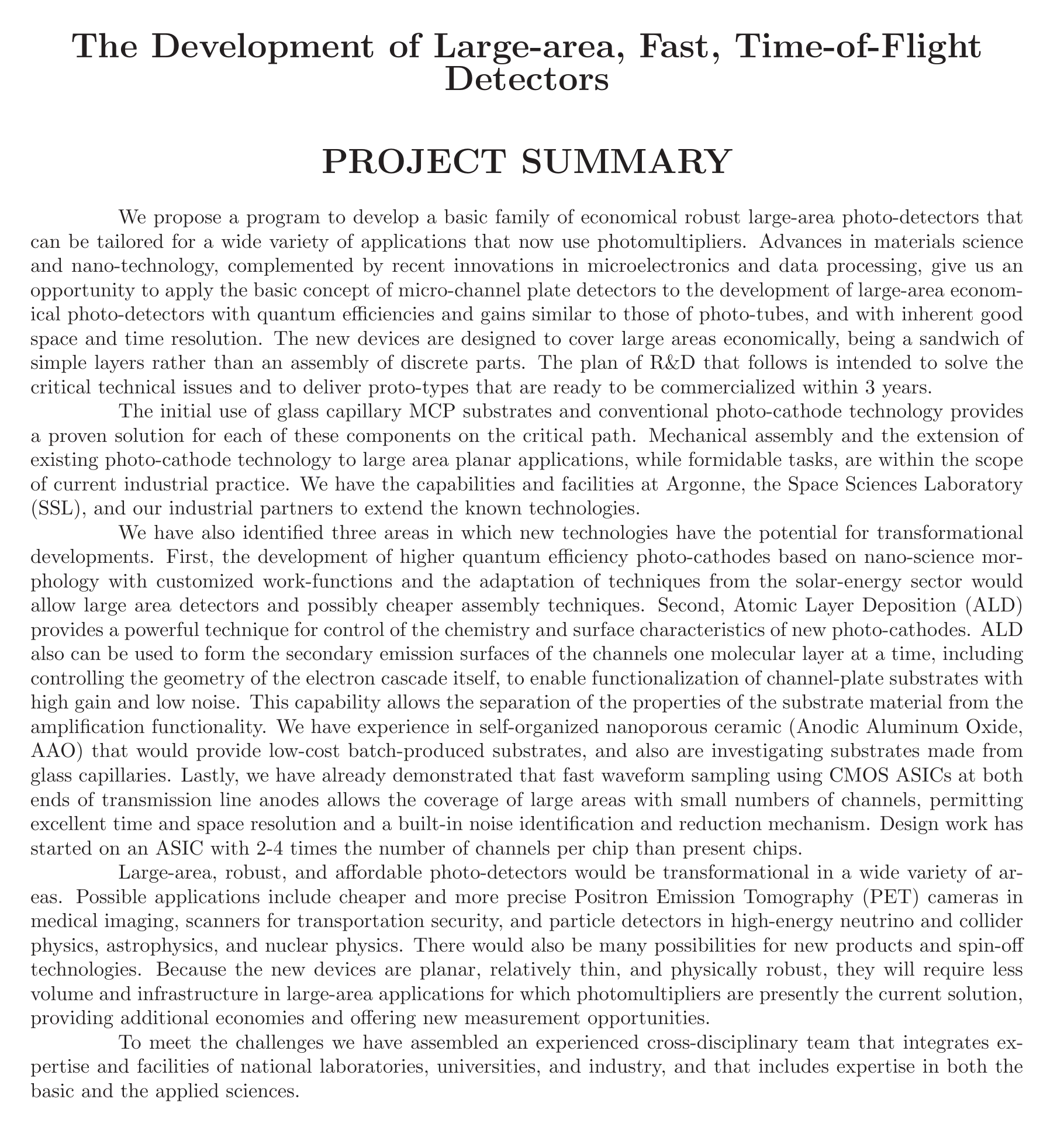}
\caption{The Project Summary page of the 2009 Proposal to the
DOE.}
\label{fig:project_summary}
\end{figure}

Figure~\ref{fig:project_summary} is the project summary page from
the proposal submission. The three areas singled out for emphasis
are: 1) high quantum efficiency  photocathode development;
2) coating of the plates by \ALD  to produce high-gain low-noise
micro-channel plates; and 3) the development of custom fast low-power waveform
sampling integrated circuits coupled to transmission lines for readout of
micro-channel-based photomultipliers.

%
%


\begin{thebibliography}{99}
\bibitem{LAPPDTM} The University of Chicago has trademarked the
LAPPD 8''-square module, with licensing to Incom Inc., to distinguish it
from small-format MCP-PMTs or fundamentally different designs.
\bibitem{Ossy_Planacon} O. Siegmund was a designer of the Planacon;
this is one of a number of areas where prior expertise was brought
directly into LAPPD.
\bibitem{NDA} The details of the proposal are covered by a
Non-Disclosure Agreement.
\end{thebibliography}
\end{document}